\documentclass[]{aa}
\usepackage{txfonts} 
\usepackage{graphicx}
\usepackage{natbib}
\bibpunct{(}{)}{;}{a}{}{,} % to follow the A&A style

\newcommand{\planck}{{\it Planck}}
\newcommand{\xmm}{{\it XMM-Newton}}

\begin{document}

\title{From universal profiles to universal scaling laws \\ in X-ray galaxy clusters}
%\subtitle{Quantifying the deviations from self-similarity}
\titlerunning{From universal profiles to universal X-ray scaling laws}

\author{S. Ettori$^{1,2}$, L. Lovisari$^{1,3}$, M. Sereno$^{1,2}$}

\authorrunning{S. Ettori et al.}

\institute{
INAF, Osservatorio di Astrofisica e Scienza dello Spazio, via Pietro Gobetti 93/3, 40129 Bologna, Italy \and
INFN, Sezione di Bologna, viale Berti Pichat 6/2, I-40127 Bologna, Italy \and
Center for Astrophysics $|$ Harvard $\&$ Smithsonian, 60 Garden Street, Cambridge, MA 02138, USA
  }
\mail{stefano.ettori@inaf.it}

\abstract
%%%  [~/DATA/OUTSKIRTS/icm_cmz] icm_cmz_sb.idl
{As the end products of the hierarchical process of cosmic structure formation, galaxy clusters 
present some predictable properties, like those mostly driven by   gravity, and some others more affected
by astrophysical dissipative processes that can be recovered from observations and 
that show remarkable universal behaviour once rescaled by halo mass and redshift.
However, a consistent picture that links these universal radial profiles and the integrated values of the thermodynamical quantities
of the intracluster medium, also quantifying the deviations from the standard self-similar gravity-driven scenario, 
has to be demonstrated.
In this work we use a semi-analytic model based on a universal pressure profile in hydrostatic equilibrium within a cold dark matter halo 
with a defined relation between mass and concentration to reconstruct the scaling laws between the X-ray properties of galaxy clusters.
We also quantify any deviation from the self-similar predictions 
in terms of temperature dependence of a few physical quantities such as the gas mass fraction, 
the relation between spectroscopic temperature and its global value, and, if present, the hydrostatic mass bias.
This model allows us to reconstruct both the observed profiles and the scaling laws between integrated quantities. 
We use the \planck\ Early Sunyaev-Zeldovich (ESZ) sample, a \planck-selected sample of objects homogeneously analysed in X-rays, to calibrate 
the predicted scaling laws between gas mass, temperature, luminosity, and total mass.
% and to guide our interpretation of the observed deviations from the standard self-similar predictions.
Our universal model reproduces well the observed thermodynamic properties and provides a way to interpret the observed deviations 
from the standard self-similar behaviour,  
% in terms of a dependence upon the temperature of (i) the ratio between the value of the temperature profile 
% and its global value, and (ii) the gas mass fraction, 
 also allowing us to define a framework to modify accordingly the characteristic physical quantities that renormalise the observed profiles.  
By combining these results with the constraints on the observed $Y_{SZ}-T$ relation we show how we can quantify the level of gas clumping 
affecting the studied sample, estimate the clumping-free gas mass fraction, and suggest the average level of hydrostatic bias present.}


\maketitle % Insert title

\section{Introduction}

\begin{table*}[ht]
\centering
\caption{Characteristic physical scales at two typical overdensities ($\Delta=500$ and $200$ times the critical density at redshift $z$) 
for an input mass  of $10^{15} M_{\odot}$, a gas mass fraction $f_{\rm gas} =0.1$, and redshift 0.
The dependence on mass and redshift are indicated in column 2; the dependence on $f_{\rm gas}$ in column 3.}
\begin{tabular}{ccccc}
\hline
Quantity  &  $f(M, z)$  &  $f(f_{\rm gas})$ & $\Delta=500$ &  $\Delta=200$ \\
\hline 
\rule{0pt}{2.5ex} $T_{\Delta}$ (keV) & $M^{2/3} \, E_z^{2/3}$ & $f_{\rm gas}^0$ & 8.95 & 6.59 \\
\rule{0pt}{2.5ex} $\bar{n}_{e, \Delta}$ (cm$^{-3}$) & $E_z^2$ & $f_{\rm gas}^1$ & 2.37 $\times 10^{-4}$ & 9.48 $\times 10^{-5}$ \\
\rule{0pt}{2.5ex} $P_{\Delta}$ (keV \, cm$^{-3}$) &  $M^{2/3} \, E_z^{8/3}$  & $f_{\rm gas}^1$ & 2.12 $\times 10^{-3}$ & 6.25 $\times 10^{-4}$ \\
\rule{0pt}{2.5ex} $K_{\Delta}$ (keV \, cm$^{2}$) &  $M^{2/3} \, E_z^{-2/3}$  &  $f_{\rm gas}^{-2/3}$ & 2338 & 3172 \\
 \hline
\end{tabular}
\label{tab:delta}
\end{table*}

Galaxy clusters are cosmological objects that form by hierarchical aggregation of matter under the action of the gravity force.
As a consequence of this force the clusters have physical properties that scale  to the mass and redshift of the dark matter halo
\cite[e.g.][]{kaiser86,boh12,kb12}. This is true not only for their total gravitating mass, but also for observational quantities
that depend mostly on the depth of the potential well, like the temperature of the intracluster medium, which  indicates
how much most of the baryons collapsed in a dark matter halo are heated by the accretion shocks.

Where the physical processes of aggregation and collapse are dominated  by the gravity, a set of relations emerges from this self-similar scenario 
between the global quantities (i.e. integrated over the volume)  that describe the observed properties of the galaxy clusters, 
such as  gas temperature, luminosity, mass, and total mass.
However, a predictable behaviour is also expected to hold in how these quantities vary with the radial distance from the bottom of the potential well,
in particular in regions away from the core. 
In the cluster's core, the relative distribution and energetics of the baryons might be affected by feedback from     star formation 
and active galactic nuclei, as well as radiative cooling, making their distribution less predictable, but not regulated by  gravity alone.
These radial profiles are then defined as universal because they should reproduce the observed profiles of any galaxy cluster, once rescaled by 
some quantities that are proportional to the mass and the redshift of the halo.

Universal radial profiles of the electron density \citep{croston08}, gas temperature \citep{vikhlinin06,baldi12}, electron pressure 
\citep{nagai07,arnaud10}, and gas entropy \citep{pratt+10} have been obtained recently as average scaled profiles, i.e.
by rescaling the measured quantities through the expected global mean value (which affects the normalisation of the profile) 
and a characteristic physical radius (e.g. $R_{500}$, which defines the radial scale).
\cite{ghi19univ} present the most recent work in which universal radial profiles of the intracluster medium (ICM) properties
are recovered out to $R_{200}$ for the X-COP sample of 12 nearby massive galaxy clusters.

In the present work we investigate how, by assuming  a universal radial profile for the gas pressure and a relation for the internal distribution 
of the cluster dark matter, we can recover the universal integrated properties of a given halo.
This semi-analytic approach is then compared with very recent observational results on the radial profiles and the scaling relations between 
hydrostatic masses, gas masses, gas luminosities, and temperatures obtained for a \planck-selected sample in \cite{lovisari20}.

The paper is organised as follows. 
In Section~2 we present our assumptions and the predictions for the universal profiles of the thermodynamic quantities.
The integrated quantities are described in Section~3, where we also discuss an application of the relations presented in \cite{ettori15}
that account for the deviations from the self-similar scenario in a physical consistent framework.
We summarise our main findings and draw our conclusions in Section~4.

In the following analysis we refer to radii, $R_{\Delta}$, and masses,  $M_{\Delta}$, which are the corresponding values 
estimated at the given overdensity $\Delta$ as $M_{\Delta} = 4/3 \, \pi \, \Delta \, \rho_{\rm c,z} R_{\Delta}^3$, 
where $\rho_{\rm c,z} = 3 H_z^2 / (8 \pi G)$ is the critical density of the universe at the observed redshift $z$ of the cluster, 
$G$ is the universal gravitational constant, and $H_z = H_0 \, \left[\Omega_{\Lambda} +\Omega_{\rm m}(1+z)^3\right]^{0.5} 
= H_0 \, E_z$ is the value of the Hubble constant at the same redshift.
For the $\Lambda CDM$ model we adopt the cosmological parameters
$H_0=70$ km s$^{-1}$ Mpc$^{-1}$ and $\Omega_{\rm m} = 1 - \Omega_{\Lambda}=0.3$.

Historically, the characteristic thermodynamic quantities within a given overdensity $\Delta$ 
refer to a singular isothermal sphere with mass $M_{\Delta}$ and radius $R_{\Delta}$ in hydrostatic equilibrium.
In this case mass and gas temperature are simply related by the equation \citep[e.g.][]{voit+05} 
\begin{equation}
\frac{G\, M_{\Delta} \, \mu m_a}{2 R_{\Delta}} = k_B T_{\Delta},
\end{equation}
where $\mu$ is the mean molecular weight of the gas, 
$m_a$ is the atomic mass unit of  $1.66 \times 10^{-24}$ g, and $k_B$ is the Boltzmann constant. 
Associating   a mean electron density $\bar{n}_{e, \Delta} = \Delta \, f_{\rm gas} \, \rho_{\rm c,z} / (\mu_e m_a)
= \mu/\mu_e \, \bar{n}_{\rm gas}$, where $f_{\rm gas} = M_{\rm gas}(<R_{\Delta})/M_{\Delta}$ is the mass gas fraction, 
$\bar{n}_{\rm gas}$ is the mean gas density, and $\mu/ \mu_e = 0.52$, with $\mu_e=1.17$ and $\mu=0.61$ 
being respectively the mean molecular weight of electrons and of the gas for a plasma with 0.3 solar abundance tabulated 
in \cite{ag89}\footnote{$\mu_e=1.16$ and $\mu=0.60$ are also obtained for the popular abundance table adopted for the X-ray spectral analysis  in \cite{aspl09}.}, we can also write the mean values for pressure,
$P_{\Delta} = k_B T_{\Delta} \, \bar{n}_{e, \Delta}$, and for entropy, $K_{\Delta} = k_B T_{\Delta}  \, \bar{n}_{e, \Delta}^{-2/3}$.
Some characteristic values of these mean physical scales are quoted in Table~\ref{tab:delta} with their dependences
on mass, redshift, and gas mass fraction.

\section{Universal thermodynamic radial profiles}

\begin{figure*}[ht]
\centering
\hbox{
\includegraphics[page=4,trim=0 40 0 230,clip,width=0.5\hsize]{e15l20_fig.pdf} 
\includegraphics[page=3,trim=0 40 0 230,clip,width=0.5\hsize]{e15l20_fig.pdf} 
}\hbox{
\includegraphics[page=2,trim=0 40 0 230,clip,width=0.5\hsize]{e15l20_fig.pdf} 
\includegraphics[page=1,trim=0 40 0 230,clip,width=0.5\hsize]{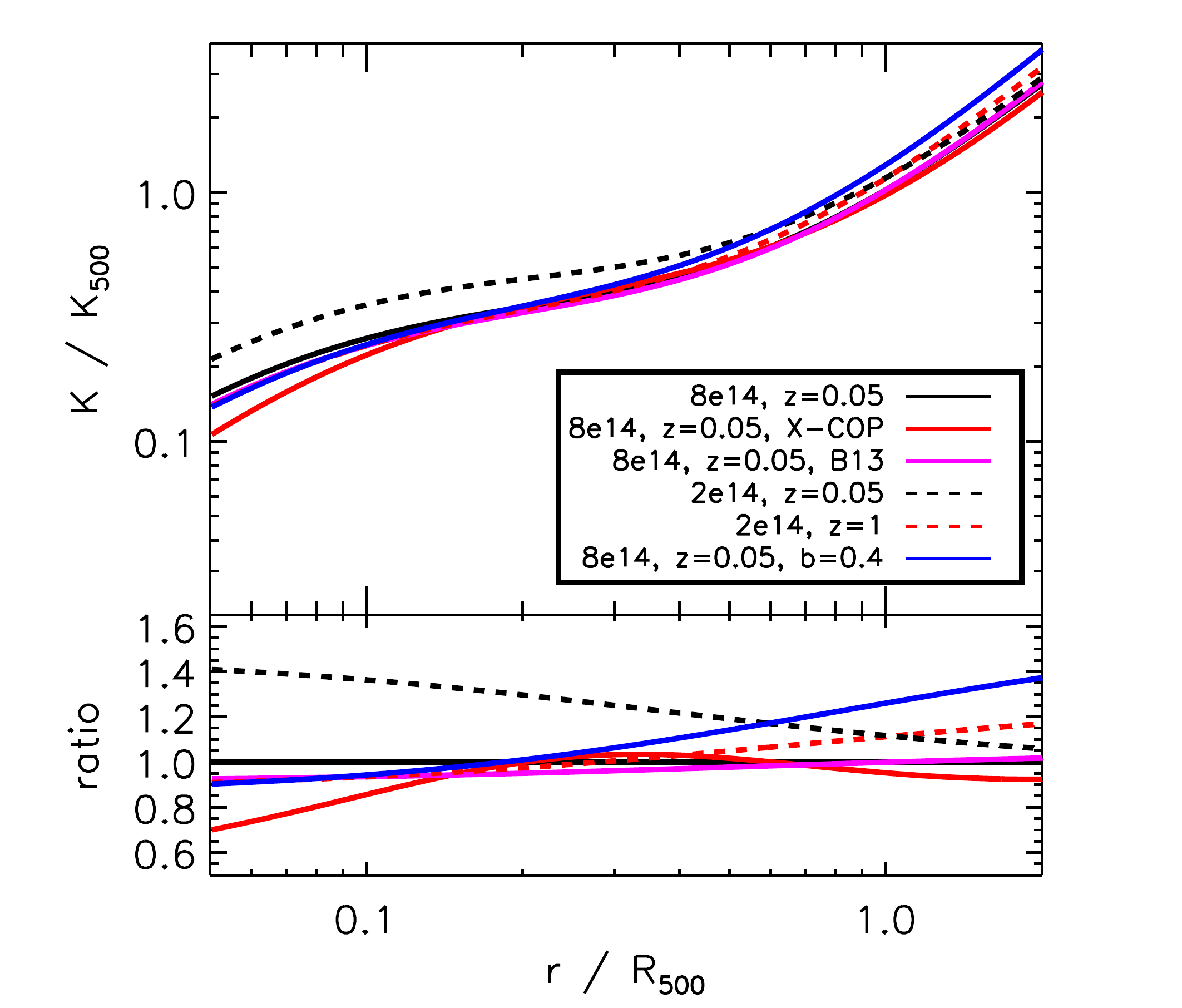} 
}\caption{ % analytical models ({\it icm\_cmz.pro \& sb\_univ.pro}) 
Reconstructed thermodynamic radial profiles for objects with $(M_{500}, z) = (8 \times 10^{14} M_{\odot}, 0.05)$,  black solid line;  
the same, but using  a pressure profile from X-COP, red solid line,  the B13 $c-M-z$ relation, purple solid line; or 
assuming a hydrostatic bias $b=0.4$, blue solid line;  $(M_{500}, z) = (2 \times 10^{14} M_{\odot}, 0.05),$  black dashed line;  
$(2 \times 10^{14} M_{\odot}, 1),$  red dashed line.
The ratio is shown with respect to the reference case, black solid line, $M_{500} = 8 \times 10^{14} M_{\odot}$ and $z=0.05$.
$P_{500}, n_{500}, T_{500}, K_{500}$ refer to the normalisation values presented in Table~\ref{tab:delta}.
} \label{fig:prof}
\end{figure*}

In this section we define the functional forms of the thermodynamic profiles for the ICM
that will be integrated to recover the global properties 
to be compared with the corresponding quantities measured in X-ray observations.
To define these functional forms, we assume an ICM in hydrostatic equilibrium within a spherically symmetric dark matter potential.
Two simple ingredients are needed:  a radial distribution for the mass, and  a radial profile for one of the 
thermodynamic quantities (gas density,  temperature,  pressure, or entropy).

To model the mass distribution we adopt a halo concentration--mass--redshift relation, $c$-$M$-$z$, 
for mass concentration $c_{200}$, global mass value $M_{200} = 4/3 \pi \, 200 \rho_{c, z} R_{200}^3$, 
and redshift of the observed object for a Navarro-Frenk-White (NFW) mass density profile 
\citep[][]{nfw97} with concentration and radius at $\Delta=200$
related through the scale radius $r_{\rm s}$: $c_{200}$ = $R_{200} /r_{\rm s}$. 
We consider the $c$-$M$-$z$ relation described in \cite{dutton14} (hereafter D14), 
with $\log c_{200} = A +B \log (M_{200} / 10^{12} M_{\odot} h_{100}^{-1})$,
$B = -0.101 +0.026 \, z$, and $A = 0.520 +(0.905-0.520) \exp(-0.617 z^{1.21})$.
We   also use  an alternative relation following the prescriptions described in \cite{bha13} (hereafter B13),
which provides lower values of concentration by $\sim$10\%\ (20\%)  
at $z\sim0.05 \, (1)$ in the mass range considered in the present study 
\citep[$10^{14}-10^{15} M_{\odot}$; see also ][for a detailed comparison between different models as a function of mass and redshift]{dk15}.
 We note that the quoted relations depend on the assumed cosmological parameters, but less significantly at $M>10^{14} M_{\odot}$, where
the differences in predicted concentrations are on the order of a few per cent for the values of $H_0$ and $\Omega_{\rm m}$ adopted here
(see e.g. figure~9 in D14). 
We also note that the quoted $c$-$M$-$z$ relations refer to the results extracted from dark matter only simulations.
It is known that the distribution of baryons in the cluster halo can affect how concentration and total mass relate, 
causing for instance a steepening of the relation because star formation is fractionally more efficient in low-mass objects, 
with an overall larger normalisation because  this effect is non-vanishing at all masses \citep[see e.g.][]{fedeli12}.
However, it has been proved that, once a proper selection of N-body simulated systems is done to mimic an observational sample,
the predicted $c$-$M$-$z$ relation matches the observed one at least for very massive objects
\citep[at the 90\% confidence level in the CLASH sample, where the total masses were recovered from the gravitational lensing signal; see][]{mer+al15}.
Furthermore, considering that the adopted $c$-$M$-$z$ relations were also used to infer the total masses in the ESZ sample 
that will be analysed in our work, and that introducing the self-gravity due to the gas (the dominant baryonic component, 
although accounting for less than 15\% of the total mass) would complicate,
without  much benefit, the calculations presented below, we use the quoted $c$-$M$-$z$ relations as input
to define the total mass of our systems.

It is worth noting that we are interested here in the global average behaviour of the cluster properties,
and do not propagate any error and/or scatter on the mean quantities.
The $c$-$M$-$z$ relation is known to have an intrinsic scatter
of $\sigma_{\log c} \sim 0.16$ on the predicted values of the concentration for given mass 
\citep[e.g.][]{dk15} that might be propagated through 
the relations used in this work to evaluate its impact on the reconstructed distribution of the 
observed thermodynamic profiles and integrated quantities of the ICM. 
We postpone to a future work further discussion on the distribution of the input parameters,
although we also consider a different set of parameters (both for the $c$-$M$-$z$ relation
and for the pressure profile) depending on the dynamical state of the systems, such as
relaxed cooling core objects that are expected to have higher halo concentration and higher values
of pressure in the core (see Sect.~\ref{sect:discussion}).

\begin{figure}
\centering
\includegraphics[page=5,trim=50 30 0 230,clip,width=1.1\hsize]{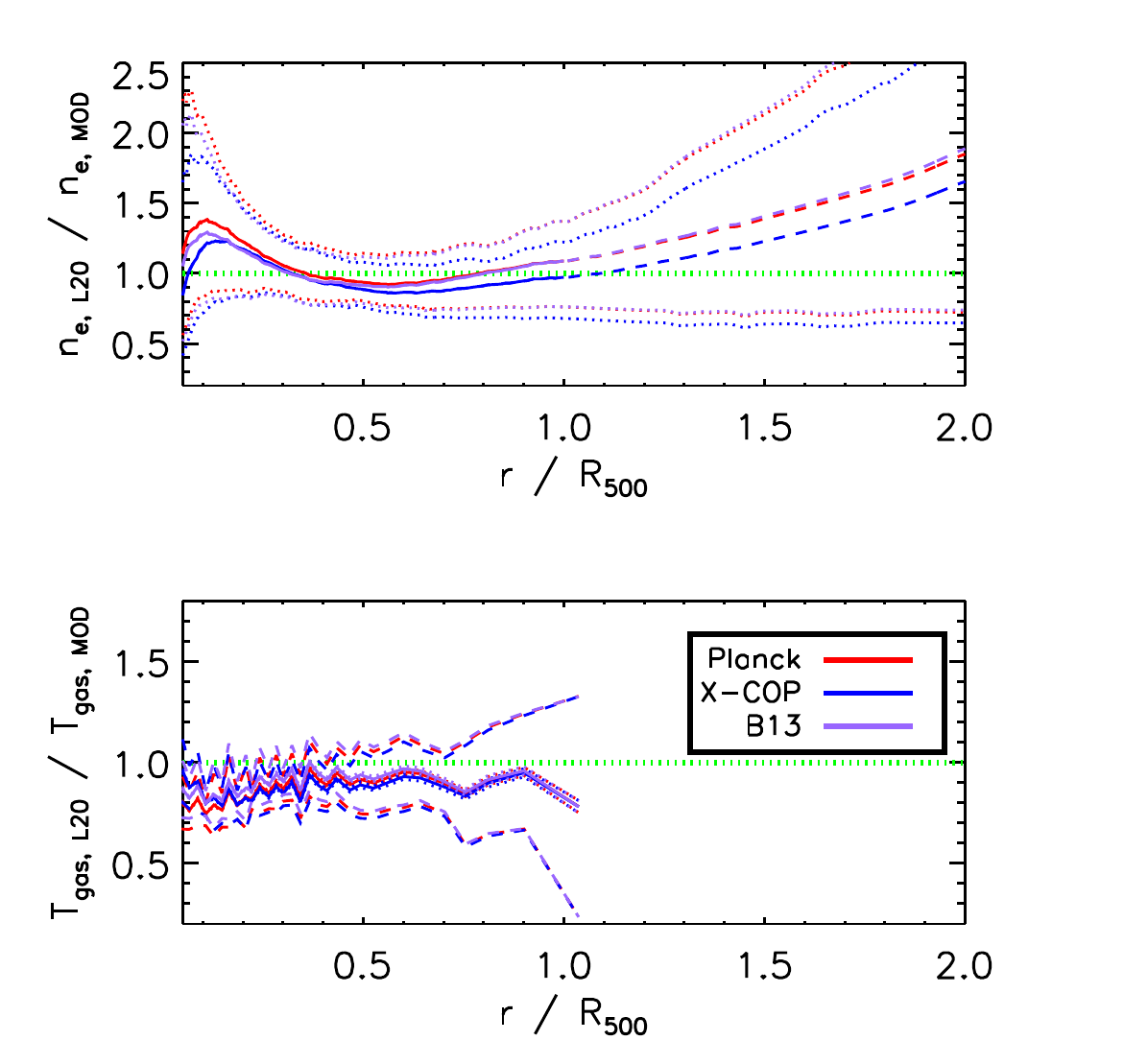} 
\caption{Comparisons between the stacked profiles from the 120 objects analysed for the ESZ sample in \cite{lovisari20}
and predictions from different models of (i) the universal pressure profile (\planck, used as reference, and
X-COP) and (ii) the $c$-$M$-$z$ relation (D14, used as reference, and B13) estimated
at the median values in the sample of $M_{500}$ and redshift.
(Top panel) Electron density profiles recovered from the best-fit parameters of a double-$\beta$ model.
The solid line indicates the median value at each radius; the dotted lines show the 16th and 84th percentile estimated at each radius.
The dashed line represents the extrapolation beyond the observational limit for the sample.
(Bottom panel) Stacked temperature profile obtained from the weighted mean of 30 spectral points
in each bin. Errors on the mean and dispersion (dotted lines) are overplotted.
} \label{fig:prof_mod}
\end{figure}

The predictions on the radial behaviour of the interesting physical quantities are then recovered from
the inversion of the hydrostatic equilibrium equation \citep[see e.g.][]{ettori+13} 
\begin{equation}
% \rho_{\rm gas} = \frac{dP_{\rm gas}}{dr} \frac{r^2}{G \, M_{\rm tot}}
n_{\rm e} = -\frac{dP_{\rm e}}{dr} \frac{r^2}{\mu \, m_a \, G \, M_{t\rm HE}}
\label{eq:mhe}
,\end{equation}
where $P_{\rm e} = n_{\rm e} T_{\rm gas}$ in units of keV cm$^{-3}$ % = \rho_{\rm gas} T_{\rm gas} / (\mu m_a) = $
is described by a generalised NFW \citep[e.g.][]{nagai07,arnaud10}
\begin{eqnarray}
P_{\rm e} & = & P_{500} \, E_z^{8/3} \, \left( \frac{M_{500}}{3 \times 10^{14} h_{70}^{-1} M_{\odot}} \right)^{\alpha_M} h_{70}^2 \, P_r  \nonumber \\
P_r & = & \frac{P_0}{(c_{500} x)^{\Gamma}\,  \left[1 + (c_{500} x)^A \right]^{(B- \Gamma) / A}}
\label{eq:pres}
\end{eqnarray}
with $x=r / R_{500}$, the normalisation $P_{500} \propto \mu/\mu_e f_{g,500} M_{500}^{2/3} = 
1.65 \times 10^{-3} \left(M_{500} / 3 \times 10^{14} h_{70}^{-1} M_{\odot} \right)^{2/3}$ keV cm$^{-3}$, \footnote{
For the sake of clarity, here $(\mu, \mu_e, f_{g,500}) = (0.59, 1.14, 0.175)$ are assumed for consistency with the original work and the associated best-fit parameters;
in the following analysis, we use the values of $\mu$ and $\mu_e$ quoted in Sect.~1.}
the parameters $(P_0, c_{500}, \Gamma, A, B)$ equal to $(6.41, 1.81, 0.31, 1.33, 4.13)$,
and $\alpha_M = 0.12$, accounting for the observed deviation from the standard self-similar scaling, set as in \cite{planck13}.
We have also considered an alternative pressure profile from the recent analysis of the joint \xmm\ and \planck\ signals of a sample
of 12 nearby massive galaxy clusters presented in \cite{ghi19univ}. Converting the published values to feed equation~\ref{eq:pres}, 
we set $(P_0, c_{500}, \Gamma, A, B)) = (5.29, 1.49, 0.43, 1.33, 4.40)$ and $\alpha_M = 0$.

 The total mass $M_{tot}$ is defined equal to $M_{\rm HE} / (1-b)$, where $M_{tot}$ is modelled with a NFW profile with parameters 
set according to the model adopted, and where the factor $(1-b)$ represents the hydrostatic bias that could affect 
the estimate of the hydrostatic masses $M_{\rm HE}$.
We note that the hydrostatic bias is propagated to the shape of the gravitating mass profile through the halo concentration, 
whereas the input value $M_{\rm HE}$ is adopted to define $M_{500}$, used for instance to estimate $P_{500}$ and $R_{500}$.
Following this procedure, we mimic the observational bias induced from the (biased) measurement of the mass on the 
normalising factor $P_{500}$ and on the definition of the region over which the physical quantities are integrated. 
Instead, other quantities that are not directly used in the analysis (e.g. $T_{500}$)  are
still defined from the bias-corrected $M_{tot}$.

We show in Fig.~\ref{fig:prof} the recovered profiles following these simple prescriptions.
We present a few cases of interest, normalised to their average quantity at $R_{500}$: 
a massive halo of $8 \times 10^{14} M_{\odot}$ at redshift 0.05, which  will be used as reference; 
the same halo with a different input pressure profile and mass bias of 0.4; a system with $1/4$ of the mass at $z=$0.05 and 1.
While the assumption of the pressure profile has a negligible impact, in particular at $r>0.1 R_{500}$, 
large deviations (on the order of $\sim 20$\%) are induced by the mass (on all the profiles) and by the bias (on the temperature and
density profile and, as a cumulative effect, on the entropy profile). The redshift causes measurable discrepancies,  but lower than $\sim20$\%.

\subsection{ESZ sample}

To test and validate the predictions of our model, we  use the radial profiles and the integrated quantities obtained 
for the ESZ sample \citep{lovisari17,lovisari20}, which contains the 120 galaxy clusters in the redshift range of $0.059<z<0.546$ 
  observed with \xmm\ and which were originally selected from the \planck\ Early Sunyaev-Zeldovich \citep[ESZ;][]{esz11} sample. 
As described in \cite{lovisari17},  these systems, which have mass and redshift distributions representing well 
the whole ESZ sample of 188 galaxy clusters, are the ones for which $R_{500}$ is completely covered by \xmm\ observations.
% allowing e.g. to estimate their morphological parameters.

As described in \cite{lovisari20},  the radial temperature profiles were  derived by 
requiring a $S/N >50$  to ensure an uncertainty of $\sim$10\% in the spectrally resolved temperature
and a source-to-background count rate ratio  higher than 0.6 to reduce the systematic uncertainties in their measurements.
On average, the temperature profiles are extracted up to $\sim R_{500}$.

The gas density profiles are presented in \cite{lovisari17}. They are recovered as the geometrical deprojection
of the best-fit results with a double-$\beta$ model of the surface brightness profile extracted from the background-subtracted 
vignetting-corrected image in the 0.3--2 keV band defined to maximise the signal-to-noise ratio.

We present in Fig.~\ref{fig:prof_mod} the comparison between the observed gas density and temperature profiles with those 
predicted from our model for a mass and redshift equal to the median values in the sample 
($M_{500} = 5.9 \times 10^{14} M_{\odot}, z=0.193$). 
Both the gas density and temperature profiles predicted from the model, the latter in particular at $r>0.3 R_{500}$, 
lie comfortably around the mean, and well within the scatter, of the observed values.

\section{Integrated quantities}
\label{sect:prof}

Using the thermodynamic profiles described in the previous section, we can reconstruct  
other derived profiles (e.g. the gas mass and the hydrostatic mass) and 
the integrated quantities (e.g.  global temperature and luminosity).

The self-similar scenario predicts properties of the ICM that depend on the mass (and redshift) of the halo \citep[see e.g.][]{voit05}.
In \cite{ettori13,ettori15} (hereafter E15) we describe the standard self-similar scenario and its possible deviations 
in terms of physical quantities depending explicitly on the mass.
However, the reconstruction of the total mass in clusters is still  affected by uncertainties \citep[see e.g.][]{pratt19} 
that make its use as a variable of reference a possible source of systematic errors.
Therefore, in this study we prefer to write the scaling relations with respect to the observed temperature $T_{\rm spec}$, 
which is a direct X-ray observable, also independent from cosmology.
For our analysis, $T_{\rm spec}$ is the gas temperature measured as an integrated spectrum in the radial range $0.15-1 R_{500}$, 
and   estimated from the profiles derived from our model as a ``spectroscopic-like'' value \citep[see][]{mazzotta+04}:
\begin{equation}
T_{\rm spec} = \frac{\int n_e(r)^2 \, T(r)^{w-0.5} \, dV(r)}{\int n_e(r)^2 \, T(r)^{w-1.5} \, dV(r)}
\label{eq:tsl}
.\end{equation}
Here $w=0.75$, $T(r) = P_e(r) / n_e(r),$ and the integrals are performed over the volume $V(r)$ of interest, either up to $R_{500}$ or
between 0.15 and 1 $R_{500}$.

The X-ray luminosity is obtained for a given gas density and temperature profile, assuming an {\tt apec} model in XSPEC with a metallicity fixed to 0.3 times the solar values
in \cite{ag89} and integrating the emissivity in cylindrical volume extending up to $3 \times R_{500}$ along the line of sight and covering an aperture between 
0.15 and 1 $R_{500}$.  

In the following we adopt the general description presented in E15.
We define the total mass $\mathcal{M} \equiv E_z M_{tot} / M_0$ ($M_0 = 5 \times 10^{14} M_{\odot}$), the gas mass $\mathcal{M}_g \equiv E_z M_g / M_{g,0}$
($M_{g,0} = 5 \times 10^{13} M_{\odot}$), and a bolometric luminosity $\mathcal{L} \equiv E_z^{-1} L / L_0$ ($L_0 = 5\times10^{44}$ erg/s). 
We write the normalisations and slopes of the scaling relations that relate these quantities to $\mathcal{T} \equiv k_B T_{\rm spec} / T_0$ ($T_0 =$ 5 keV) as
\begin{align}
\mathcal{M} & = k_M \, (1-b)^{-1} \, f_T ^{3/2} \, \mathcal{T}^{3/2}, \nonumber \\
\mathcal{M}_g & = k_M \, f_g \, f_T ^{3/2} \, \mathcal{T}^{3/2}, \nonumber \\
\mathcal{L} & = k_L \, f_g^2 \, f_T ^{3/2} \, \mathcal{T}^2,  
\label{eq:scalaw}
\end{align}
where we have defined the following parameters: 
$f_T = T(R_{500}) / T_{500} \times T_{500} / T_{\rm spec}$ relates the gas temperature at $R_{500}$ 
with the observed global value $T_{\rm spec}$; 
the X-ray--measured gas mass fraction $f_g = M_g / M_{\rm HE} = (1-b)^{-1} M_g / M_{tot} = 
\equiv  C^{0.5} \, f_{nc} / f_{g,0}$ is normalised to $f_{g,0}=0.1$ and is related to the clumping-free gas mass fraction $f_{nc}$ through
the clumping factor $C = <n_{\rm gas}^2> / <n_{\rm gas}>^2$, defined
as the ratio of the average squared gas density to the square of the mean gas density and that
affects the measurement of the gas density as obtained from the deprojection of the X-ray free-free emission 
\citep[e.g.][]{nagai+11,roncarelli+13,vazza+13,eckert+15};
$k_M$ and $k_L$ (here referring to the case of a bolometric luminosity integrated over a spherical volume up to $R_{500}$)
are defined by constants (e.g. the overdensity $\Delta$) and parameters that describe the shape of the thermodynamic profiles
and are fully described in appendix~\ref{app:km_kl}.
We note that the true gas mass fraction, i.e. the  unbiased value from the hydrostatic bias and the clumping factor, is 
written in our notation as $f_{\rm gas} = M_g / M_{tot} = (1-b) \, f_g = (1-b) \,  C^{0.5} \, f_{nc}$.

\begin{figure}
\includegraphics[page=6,trim=10 35 30 240,clip,width=\hsize]{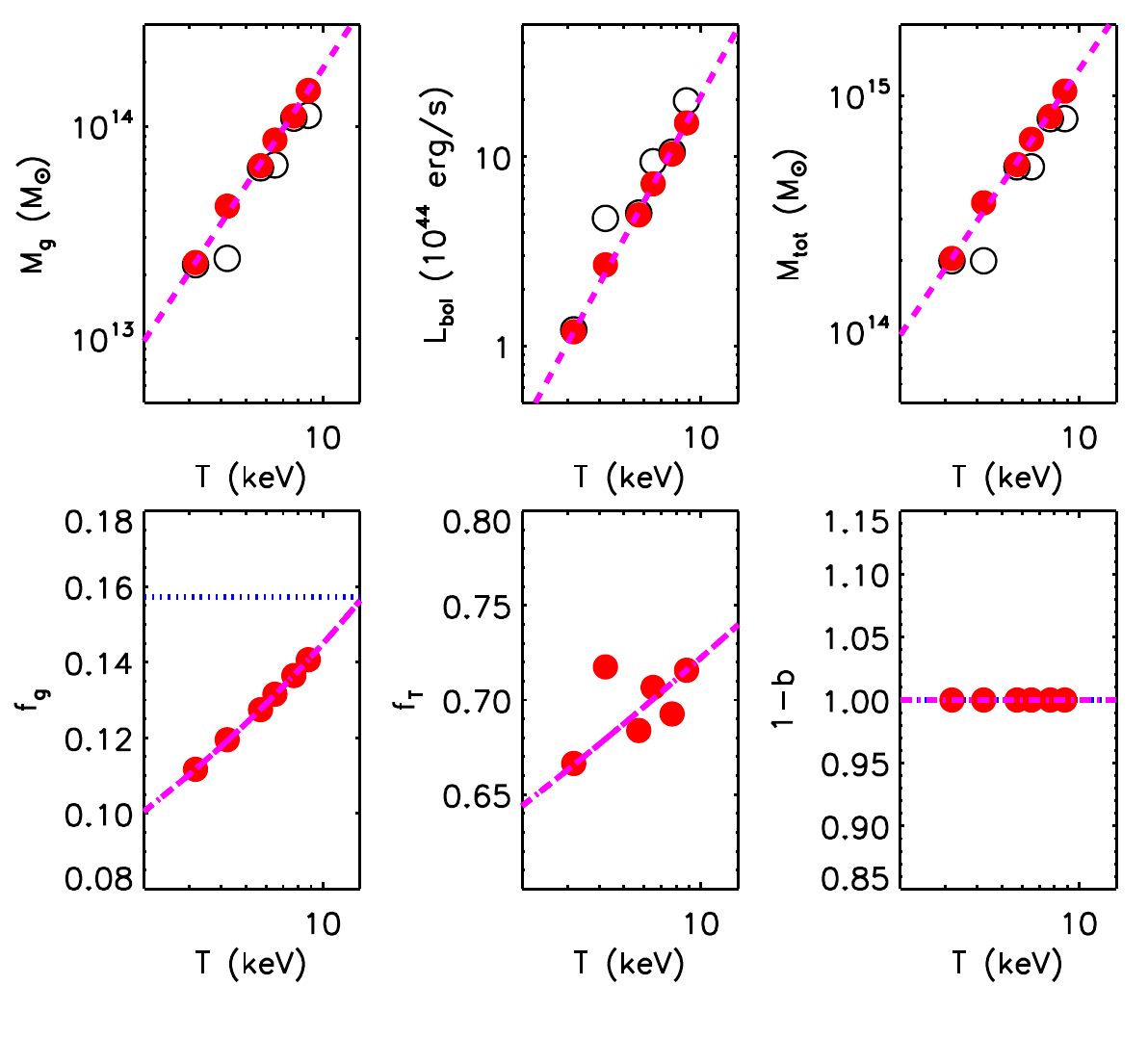}
\caption{(Top panels) Global properties (empty black dots) and  after correction for the $E_z$ factor (solid red
dots) 
recovered from their thermodynamic profiles for the following input values for $(M_{500}, z)$: 
($8 \times 10^{14} M_{\odot}, 0.05$), ($5 \times 10^{14} M_{\odot}, 0.05$), ($2 \times 10^{14} M_{\odot}, 0.05$),
($8 \times 10^{14} M_{\odot}, 0.5$), ($5 \times 10^{14} M_{\odot}, 0.5$), ($2 \times 10^{14} M_{\odot}, 1$).
The dashed magenta lines identify the best-fit relations of Eq.~\ref{eq:scalaw_t}.
A core-excised bolometric luminosity is considered.
(Bottom panels) Red dots are the quantities (from left to right:
$f_g$, $f_T$, $b$) estimated in our model
for the following input values for $(M_{500}, z)$: 
($8 \times 10^{14} M_{\odot}, 0.05$), ($5 \times 10^{14} M_{\odot}, 0.05$), ($2 \times 10^{14} M_{\odot}, 0.05$),
($8 \times 10^{14} M_{\odot}, 0.5$), ($5 \times 10^{14} M_{\odot}, 0.5$), and ($2 \times 10^{14} M_{\odot}, 1$);
the blue dotted line indicates $\Omega_b / \Omega_m$=0.157 \citep{planck+16}; the 
dashed magenta lines show the predictions following the relations:
$f_g = 0.124 \, \mathcal{T}^{0.23}$, $f_T = 0.687 \, \mathcal{T}^{0.07}$ and $1-b=h_0=1$,
where $f_g$ and $f_T$ are  the absolute values, i.e. multiplied by the assumed normalisations
of 0.1 and $\bar{f_T} = 0.71$, respectively.
} \label{fig:model}
\end{figure}

As extensively discussed in E15, any deviation from the standard self-similar behaviour can be ascribed to processes that impact 
the relative distribution of the gas and that, in the present study, we assume to be described as a power-law dependence on the gas temperature 
measured in the spectroscopic analysis $\mathcal{T} \equiv k_B T_{\rm spec} / 5$ keV:
\begin{align}
f_g = f_0 \mathcal{T}^{f_1}, \nonumber \\
f_T = t_0 \mathcal{T}^{t_1}, \nonumber \\
(1-b) = h_0.
\label{eq:func_t}
\end{align}
Here $f_0$ is chosen to be 0.1 and is  roughly representative of measured gas mass fractions, 
while $t_0$ is set to $\bar{f_T}=0.71$, the median value for our model as derived in Appendix~\ref{app:km_kl}.
We list these parameters, with their descriptions and definitions, in Table~\ref{tab:mod}.

\begin{table*}[ht]
\centering
\caption{Descriptions and definitions of the parameters used in our model. 
We define $\mathcal{T} \equiv k_B T_{\rm spec} / 5$ keV.}
\setlength{\tabcolsep}{0.8pt}
\begin{tabular}{ccc}
\hline
Quantity  &  Description & Definition \\
\hline 
\rule{0pt}{2.5ex} $(1-b) = h_0$ & hydrostatic bias & $(1-b) = M_{\rm HE} / M_{tot}$ (see Eq.~\ref{eq:mhyd}) \\
\rule{0pt}{2.5ex} $f_T = t_0 \mathcal{T}^{t_1}$ & gas temperature at $R_{500}$ normalised to the global $T_{\rm spec}$ & $f_T = T(R_{500})/ T_{\rm spec}$ \\
\rule{0pt}{2.5ex} $f_g = f_0 \mathcal{T}^{f_1}$ & X-ray-measured gas mass fraction & $f_g = M_g / M_{\rm HE}$  \\
\rule{0pt}{2.5ex} $f_{nc} = f_{nc,0} \mathcal{T}^{f_{nc,1}}$ & clumping-corrected gas fraction & $f_g = C^{0.5} \, f_{nc}$ (see Eq.~\ref{eq:c0}) \\
\rule{0pt}{2.5ex} $C = C_0 \mathcal{T}^{C_1}$ & clumping factor & $C^{0.5} = f_g / f_{nc}$ \\
\rule{0pt}{2.5ex} $f_{\rm gas} = C^{-0.5} \, M_g / M_{tot}$ & true (unbiased) gas mass fraction &  $f_{\rm gas} = (1-b) \, f_g = (1-b) \,  C^{0.5} \, f_{nc}$ \\
\hline
\end{tabular}
\label{tab:mod}
\end{table*}

We insert them in Eq.~\ref{eq:scalaw} to write these equations in their full dependence on $T$ as 
\begin{align}
\mathcal{M} & = k_M \, h_0^{-1} \, t_0^{3/2} \, \mathcal{T}^{3/2 +3/2 t_1} = N_{MT} \, \mathcal{T}^{S_{MT}}, \nonumber \\
\mathcal{M} _g & = k_M \, f_0 \, t_0^{3/2} \, \mathcal{T}^{3/2 +3/2 t_1 + f_1 } = N_{GT} \, \mathcal{T}^{S_{GT}}, \nonumber \\
\mathcal{L}  & = k_L \, f_0^2 \, t_0^{3/2} \, \mathcal{T}^{2 +3/2 t_1 +2 f_1} = N_{LT} \, \mathcal{T}^{S_{LT}},
\label{eq:scalaw_t}
\end{align}
where we represent with $N$ and $S$ the best-fit estimates of normalisation and slope, respectively, 
of the corresponding scaling laws that are quoted in Table~\ref{tab:n_s}.

% In the following, we neglect the dependence upon $E_z$.
Then, by simple algebraic calculations, we can invert equations ~\ref{eq:scalaw_t} and obtain
the normalisations $f_0, t_0, h_0$ and slopes $f_1, t_1$:
\begin{align}
f_0 & = h_0^{-1} N_{GT}/N_{MT} = (N_{LT}/k_L) \, (N_{GT}/k_M)^{-1},  \nonumber \\
t_0 & = (h_0 N_{MT}/k_M)^{2/3} = (N_{GT}/k_M)^{4/3} \, (N_{LT}/k_L)^{-2/3}, \nonumber \\
h_0 & = (N_{GT}/k_M)^2 \,  (N_{MT}/k_M)^{-1} \, (N_{LT}/k_L)^{-1},  \nonumber \\
t_1 & = 2/3 S_{MT} - 1, \nonumber \\
f_1 & = S_{GT} -S_{MT}.
\label{eq:param}
\end{align}

In Fig.~\ref{fig:model}, we show the reconstructed relations with the ICM temperature of the gas mass fraction, 
gas mass, and bolometric luminosity, and the corresponding best-fit relations that allow us to evaluate 
the normalisations $N$ and slopes $S$. These values are then used to check the predicted behaviour 
of some intrinsic properties (e.g. the dependence on $T$ of the gas mass fraction,
and of the ratio $T(R_{500})/T$; see panels at the bottom of Fig.~\ref{fig:model}).

It is worth noting  that, in order to compare our predictions with robust estimates of the X-ray luminosity, 
we integrate the bolometric luminosity within the radial range $0.15-1 R_{500}$, using quantities 
projected along the line of sight to mimic the observational values.
These characteristics of the luminosity L (i.e. if projected or integrated over spherical shells, the radial range of integration, 
the energy band) define the proper value of the constant $k_L$.
Given the constraints on the $L-T$ relation, we can invert it to recover $k_L$ under the assumption that there is no hydrostatic bias (i.e. $(1-b) = h_0 = 1)$, so that $t_0 = (N_{MT}/k_M)^{2/3}$, $f_0 = N_{GT} / N_{MT}$, and 
$k_L = N_{LT} f_0^{-2} t_0^{-3/2} = 0.503$.
We adopt this value of $k_L$ in the following analysis.

\subsection{Calibration of the model in E15}

Our semi-analytic model makes use only of  a universal pressure profile and  a $c$-$M$-$z$ relation, 
under the assumption that the total mass profile is described by a NFW model.
The integrated quantities predicted from this model are used to calibrate the relations presented in E15
and described by  Equations~\ref{eq:scalaw}--\ref{eq:param}.
By fitting in a robust way the distribution of points plotted in Fig.~\ref{fig:model} when $b$ is assumed to be zero, 
we obtain from Eq.~\ref{eq:param} that these integrated quantities relate between them following the scaling laws 
built from self-similar relations and modified by including the following dependence of the temperature profile $f_T$ 
and of the gas mass fraction $f_g = C^{0.5} f_{nc}$ on $T_{\rm spec, 0.15-1 R_{500}}$:
\begin{equation}
\begin{cases} 
f_T = 0.687 \, (0.711) \times (T/5 {\rm keV})^{0.07 (0.10)} \\
f_g = 0.124 \, (0.128) \times (T/5 {\rm keV})^{0.23 (0.25)}
\end{cases}.
\label{eq:mod}
\end{equation}
Here we quote in parentheses the results obtained by assuming a B13 model (where D14 is our model of reference) 
for the $c$-$M$-$z$ relation.

\begin{figure}[ht]
\includegraphics[page=8,trim=0 40 0 230,clip,width=\hsize]{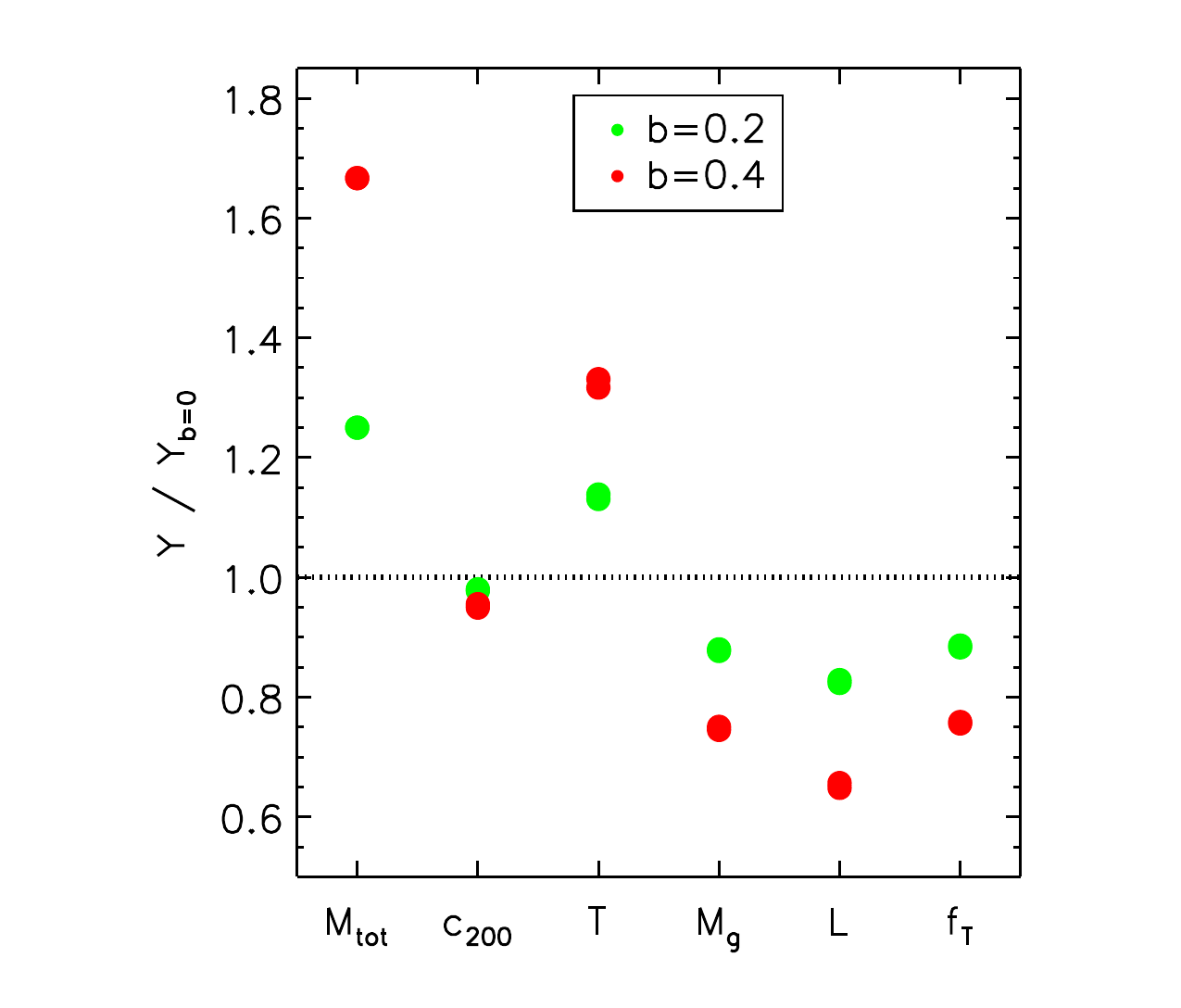} 
\caption{
Impact of the hydrostatic bias $b$ on the integrated quantities,
represented as the ratio between the quantities estimated for an assigned bias, and with bias equal to zero.
From left to right: this ratio for total mass, temperature, gas mass, core-excised bolometric luminosity,
and $f_T = T(R_{500}) / T_{\rm spec}$.
% and the parameters describing $f_T = t_0 \mathcal{T}^{t_1}$ and $f_g = f_0  \mathcal{T}^{f_1}$ (slopes indicated by empty dots).
Shown  is the case of a system with $M_{500} = 5 \times 10^{14} M_{\odot}$ at $z=$0.05 and 0.5
(the values at different redshifts overlap for most of the quantities).
} \label{fig:bias}
\end{figure}

\subsection{Role of the hydrostatic bias}
\label{sect:bias}

As described above, the hydrostatic bias is propagated to the total mass as $M_{tot} = M_{\rm HE} / (1-b)$, 
where $M_{tot}$ is modelled with the adopted NFW profile, and the hydrostatic bias modifies the shape of the gravitating mass profile 
through the halo concentration. This affects all the thermodynamic quantities depending on the dark matter distribution, (e.g.
$T_{500}$) that is defined from the bias-corrected $M_{tot}$.
On the other hand, the input value $M_{\rm HE}$ is used to rescale the observed properties (e.g. $R_{500}$ and $P_{500}$)
to mimic the observational bias once a biased $M_{\rm HE}$ is measured instead of the true value $M_{tot}$.

In Fig.~\ref{fig:bias}, we represent the impact of the hydrostatic bias as the ratio between quantities estimated with a given $b$ value and with $b=0$.
A clear trend is present, with higher biases producing higher global temperatures, and lower estimates of $M_g$ and $L$, mostly as
consequence of the increased halo mass, associated with the corresponding reduction in the halo concentration and constant value of  $R_{500}$. 
We have also modelled the trend we observe in the physical quantities as a function of the assumed bias $b$ with a functional form 
\begin{equation}
Y / Y_{b=0} =  \alpha_0 \, (1-b) +(1-\alpha_0) \, (1-b)^{\alpha_1}
,\end{equation}
such that when $Y=M_{tot}$, then $(\alpha_0, \alpha_1) = (0, -1)$.
For a typical object with $M_{500} = 5 \times 10^{14} M_{\odot}$ at $z$=0.05 shown in Fig.~\ref{fig:bias},
this functional form reproduces the observed trends within 2\% and provides the following best-fit parameters $(\alpha_0, \alpha_1)$:
 $(-0.12, -0.45)$ for $T_{\rm spec}$; $(0.34,0.39)$ for the gas mass; $(0.83, 0.25)$ for the bolometric luminosity; $(0.012, 2.61)$ for the gas fraction.
 Very similar results are obtained for systems at lower masses and higher redshifts on $T_{\rm spec}$, $M_g$, and gas mass fraction.

This translates into the following representation in the E15 formalism for the two cases $b=0.2$ and $b=0.4$
(the latter in brackets), assuming the same set of input values of $(M_{500}, z)$
\begin{equation}
\begin{cases} 
f_T = 0.597 \, (0.500) \times (T/5 {\rm keV})^{0.08 (0.10)} \\
f_g = 0.106 \, (0.087) \times (T/5 {\rm keV})^{0.24 (0.25)},
\end{cases}
\label{eq:mod_bias}
\end{equation}
with lower normalisations for $f_g$ and $f_T$ for higher values of $b$, and no significant change in the dependence on the temperature.

\begin{figure*}[ht]
\centering
\hbox{
\includegraphics[page=9,trim=0 0 0 200,clip,width=0.33\hsize]{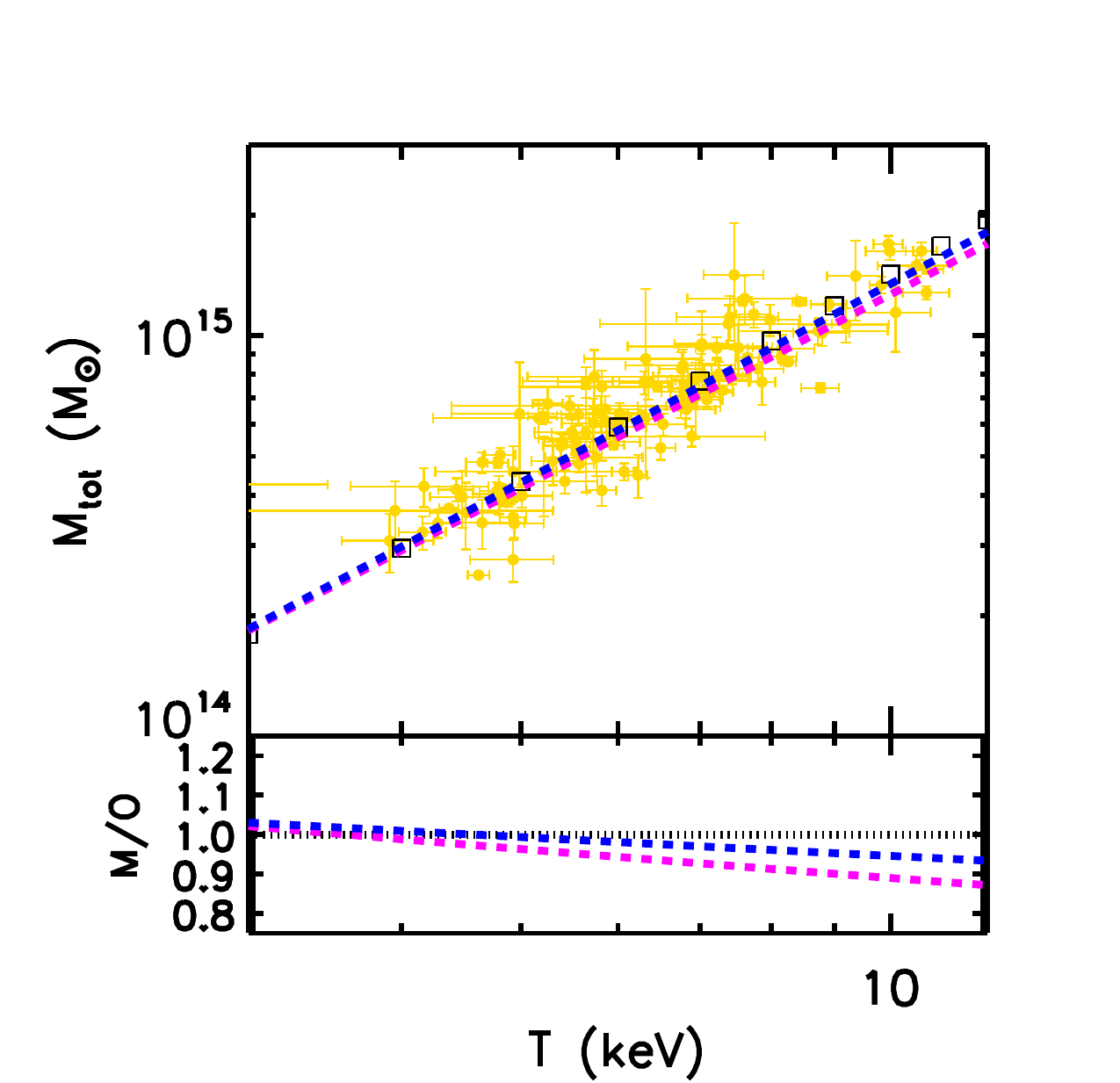}
\includegraphics[page=10,trim=0 0 0 200,clip,width=0.33\hsize]{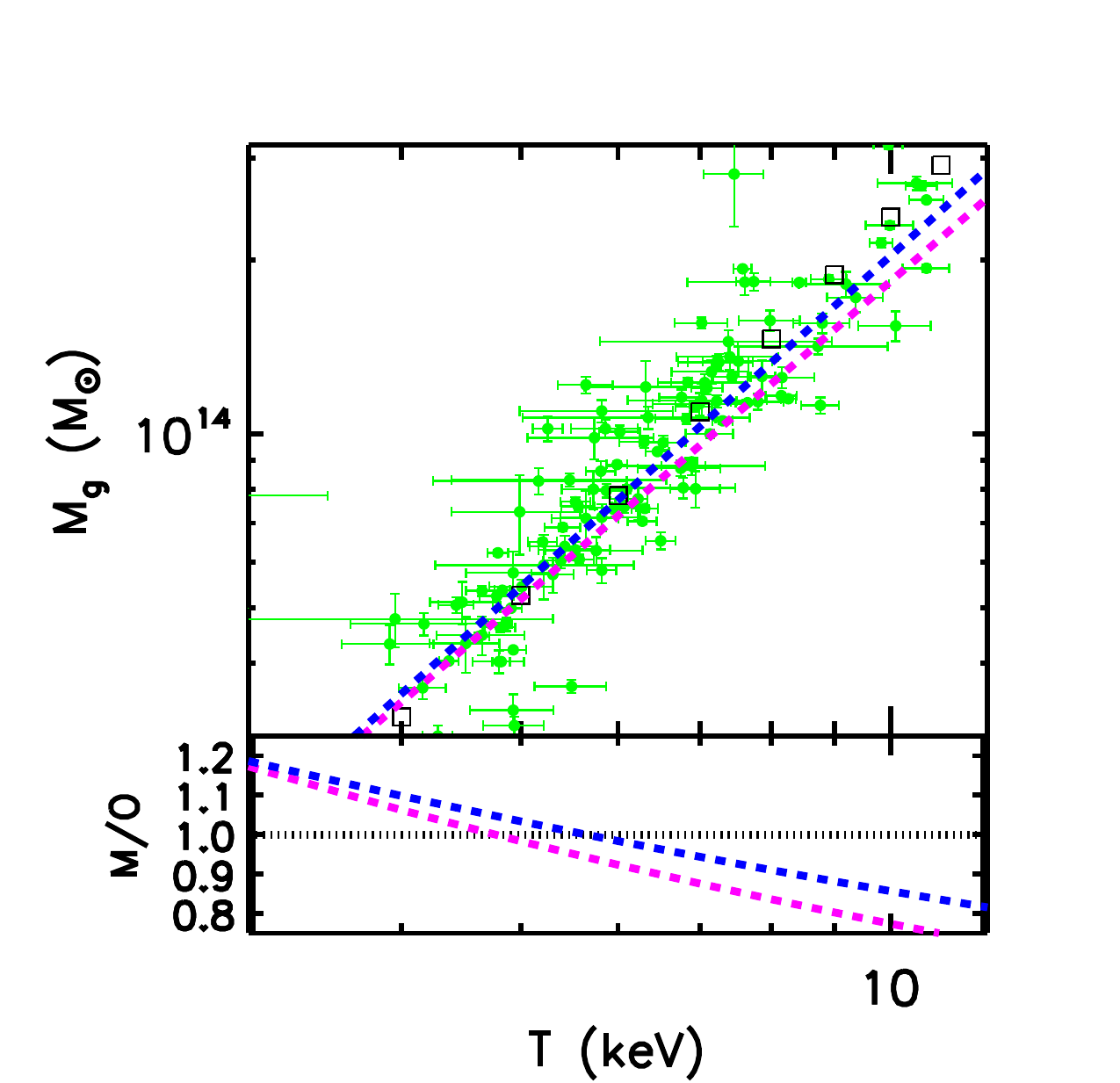}
\includegraphics[page=11,trim=0 0 0 200,clip,width=0.33\hsize]{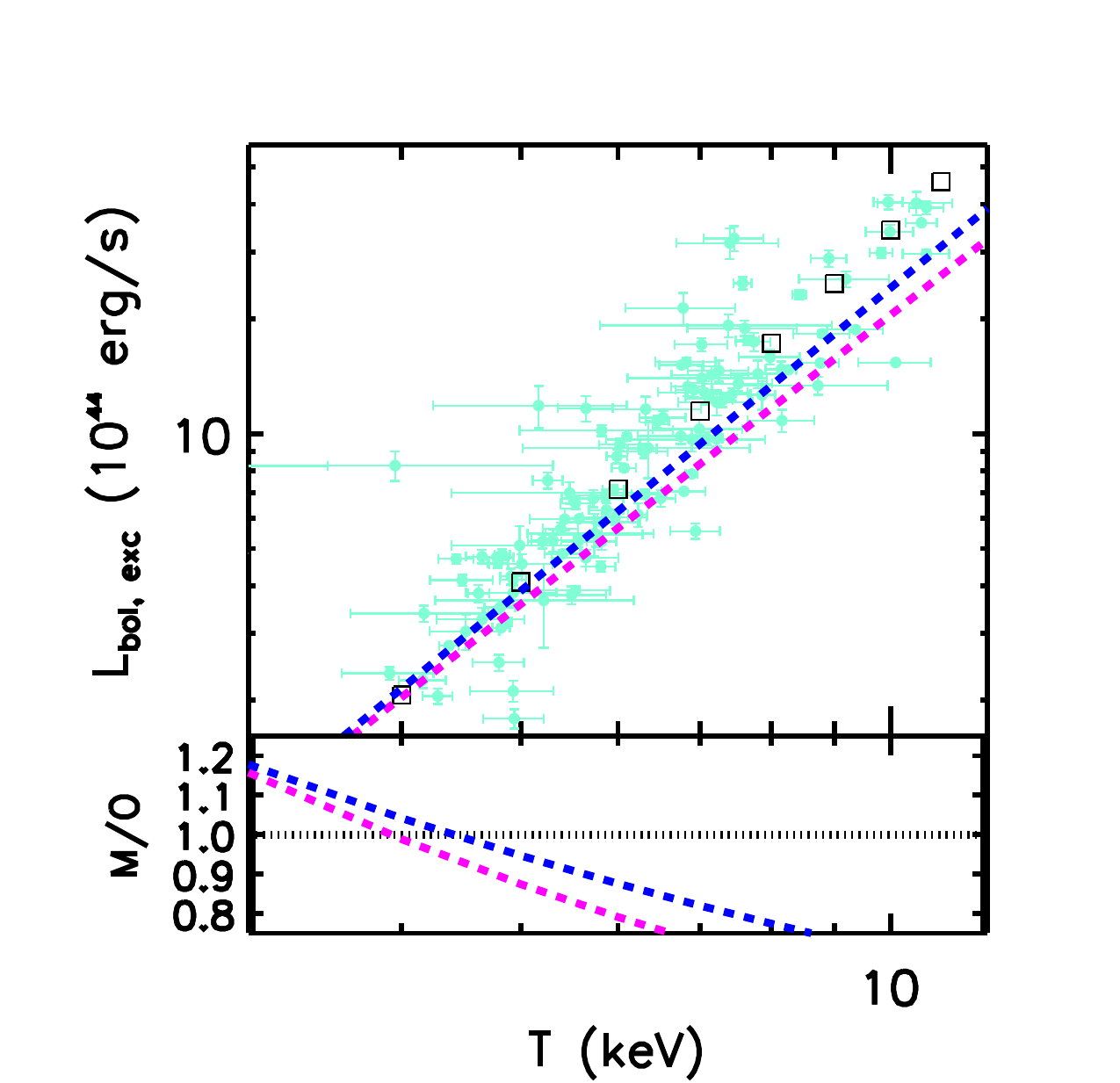}
}
\caption{
Distribution of the observed values for the ESZ sample with best-fit results from a linear fit in the logarithmic space 
\citep[using {\tt LIRA};][open squares]{ser16lira}.
Predictions from the semi-analytic model (best-fit values in Table~\ref{tab:n_s}) are overplotted 
with dashed lines in magenta (for the D14 $c$-$M$-$z$ relation) and blue (B13).
Bottom panels: Dashed lines indicate the ratios of the model (M) to 
the observed (O) best-fit relations.
} \label{fig:esz}
\end{figure*}

\begin{table}
\centering
\caption{Best-fit normalisations and slopes of the scaling laws in Eq.~\ref{eq:scalaw_t}.
We quote the constraints obtained by using a sample representative of the distribution of masses and
redshifts observed in the ESZ sample estimated with a D14 model of the $c$-$M$-$z$ relation (in parentheses, the results for a B13 model) 
and the results from  {\tt LIRA} on the ESZ sample (see Fig.~\ref{fig:esz}).}
\setlength\tabcolsep{5pt}
\begin{tabular}{ccc}
\hline
 Relation & $N$ & $S$ \\ \hline
$M-T$ & 0.83 (0.86) / 0.86$\pm$0.03 & 1.60 (1.65) / 1.72$\pm$0.10 \\
$M_g-T$ & 1.03 (1.09) / 1.05$\pm$0.04 & 1.83 (1.90) / 2.17$\pm$0.13 \\
$L-T$ & 0.71 (0.77) / 0.82$\pm$0.04 & 2.52 (2.64) / 3.07$\pm$0.18 \\
\hline
\end{tabular}
\label{tab:n_s}
\end{table}

\subsection{Results on the ESZ sample}
\label{sect:esz}

\cite{lovisari20} present a study of the X-ray scaling relations for the ESZ sample, based on hydrostatic mass profiles recovered
from the B13 $c$-$M$-$z$ relation, and using core-excised luminosities and spectroscopic temperatures, as
we reproduce in our semi-analytic model.
The best-fit relations for the observational data are corrected for the Eddington bias, 
but not for the Malmquist bias, which is negligible when fitting the X-ray properties 
of an SZ selected sample \citep[for more details, see][]{lovisari20}.
The slopes of all the investigated scaling relations are found to deviate significantly from the self-similar predictions, 
if self-similar redshift evolution is assumed. 
We  re-estimated normalisations and slopes of the scaling relations of interest using the package {\tt LIRA} \citep[][; see Table~\ref{tab:n_s}]{ser16lira}.
In Fig.~\ref{fig:esz}, we overplot these best-fit relations to the data points in the ESZ sample.

By applying our universal model, we require the self-similar predictions to be corrected 
by the following dependences on the gas temperature
\begin{equation}
\begin{cases} 
f_{T, ESZ} = 0.697 (\pm 0.103) \times (T/5 {\rm keV})^{0.15 (\pm 0.06)} \\
f_{g, ESZ} =  0.121 (\pm 0.045) \times (T/5 {\rm keV})^{0.45 (\pm 0.09)}
 \end{cases}.
 \label{eq:esz}
 \end{equation}
Here we  propagate the errors on the best-fit parameters,
and (as in the following analysis) we quote the absolute values of $f_g$ and $f_T$, 
i.e. multiplied by their normalisation values of $f_{g,0}=0.1$ and $\bar{f_T} = 0.71$.

Then  we compare the observed distribution with the predictions from our semi-analytic model.
To mimic the distribution in mass and redshift of the ESZ sample, we  simulated ten objects with our semi-analytic code
having mass and redshift equal to the median values estimated in bins of 12 clusters each, sorted in redshift. 
We overplot these predictions to the data points in Fig.~\ref{fig:esz}.
We also use a different $c$-$M$-$z$ relation to probe the dependence on the assumed model.
As previously discussed, the B13 model predicts lower $c_{200}$ at higher redshift, for a fixed halo mass.
This behaviour induces differences in the estimates of $T$, $L$, and gas mass. 
For example, a halo with $M_{500}=5 \times 10^{14} M_{\odot}$
is expected to have $c_{200}$ lower by 5\%\ and 13\% at $z=$0.05 and 0.5 in B13, 
inducing a slightly different distribution of the dark matter and consequent reshaping of the
thermodynamic profiles that produce $T$ lower by 2\%\ and 4\%, $M_g$ higher by 2\%\ and 4\% 
and $L$ higher by 3\%\ and 9\%, respectively. 

The normalisations are within the $1 \sigma$ intervals of the observational constraints 
(the largest tension being the results on $N_{LT}$ from the D14 model), whereas the slopes of the scaling laws
in the ESZ sample tend to be systematically higher than the predicted values by 0.7--1.2 $\sigma$, 2.6--2.1  $\sigma$,
and 3--2.4 $\sigma$ for the $M-T$, $M_g-T$, and $L-T$ relation with D14 and B13 models, respectively.
It is important to note, however, that the steepening of the relations might depend on the selection applied 
and on how the fit is performed \citep[in our case, we consider in {\tt LIRA} the scatter on both the variables $X$ and $Y$, 
i.e. we leave  the parameters {\tt sigma.XIZ.0} and {\tt sigma.YIZ.0} free to vary; for details, see][]{ser16lira}.
 For instance, if we fix  {\tt sigma.XIZ.0} =0, then
the relations become flatter, with slopes of 1.60, 2.00, and 2.82 for the $M-T$, $M_g-T$, and $L-T$ relation, respectively. 
In general, we obtain a better agreement between the observed distributions in the ESZ sample and the simulated data
when we apply the B13 model as a consequence of the shift to lower temperatures for a given mass, and conclude
that the distribution of the integrated physical quantities follows what is reconstructed from the universal model
within $10$\%, on average.

Using Equations~\ref{eq:param} on these simulated data, and assuming a D14 model, the best-fit values convert into
$(t_0, t_1) = (0.689, 0.07)$; $(f_0, f_1) = (0.122, 0.22)$ for $h_0$ free to vary (best-fit: 1.02) and 
$(t_0, t_1) = (0.680, 0.07)$; $(f_0, f_1) = (0.124, 0.22)$ for $h_0$ fixed to one.
When B13 is adopted we obtain $h_0=1.01$ once it is left to vary,  
$(t_0, t_1) = (0.698, 0.10)$; $(f_0, f_1) = (0.125, 0.25)$, and
$(t_0, t_1) = (0.694, 0.10)$; $(f_0, f_1) = (0.126, 0.25)$ with $h_0$ fixed to one.
Overall, there is a good consistency with the results shown in Eq.~\ref{eq:esz}, apart from a clear steeper dependence
of the gas fraction on $T$ that will be  discussed further in the next subsection.

\subsection{Including the distribution of $Y_{SZ}$}
\label{sect:yt}

The same plasma responsible for the X-ray emission can also be traced through the Sunyaev-Zeldovich (SZ) effect, 
generated from the Compton scattering of the photons of the cosmic microwave background on the electrons of the ICM \citep{sz}.
The millimetre wave emission due to the thermal SZ effect is proportional to the integrated pressure 
of the X-ray emitting plasma along the line of sight and is described, as aperture integrated signal \citep[see e.g.][]{mro19},
by the integrated Compton parameter $Y_{SZ} D_A^2 = (\sigma_T/ m_e c^2) \int P dV$, 
where $D_A$ is the angular diameter distance to the cluster, 
$\sigma_T = 8 \pi/3 ( e^2 / m_e c^2 )^2 = 6.65 \times 10^{-25}$ cm$^2$ is the Thomson cross section, 
$m_e$ and $e$ are respectively the electron rest mass and charge, $c$ is the speed of light, 
and $P = n_e T$ is the electron pressure profile.

Using the formalism presented in E15, this signal relates to the gas temperature as
\begin{equation}
\frac{E_z Y_{SZ} D_A^2}{10^{-4} {\rm Mpc}^2} = k_{SZ} \, f_T^{5/2} \, f_{nc} \, \mathcal{T}^{5/2} =  N_{YT} \, \mathcal{T}^{S_{YT}}
\label{eq:ysz_t}
,\end{equation}
where $k_{SZ}$ is derived in Appendix~\ref{app:ksz}.

It is worth noting a few issues concerning the reconstruction of the SZ signal.
First, the SZ signal depends linearly on the pressure profile and is sensitive to the upper limit of the integration. 
We collect the estimates of the spherically integrated SZ flux up to $R_{500}$
$Y_{SZ}$ from the PSZ1 catalogue\footnote{We  use
the catalogue {\tt PSZ1v2.1.fits} available, with description, at
http://szcluster-db.ias.u-psud.fr/.} \citep{PSZ1catalog}.
% https://wiki.cosmos.esa.int/planckpla2015~/index.php/~Catalogues\#Union\_catalogue.} 
%\citep{catalog2}, by inverting the equation that defines the quoted masses $M_{SZ}$:
%$Y_{SZ} D_A^2 / 10^{-4} {\rm Mpc}^2 = 10^{-0.19} \times \left(M_{SZ} / 6\times10^{14} M_{\odot} \right)^{1.79} E_z^{0.66}$
%\citep[see eq.~7 in ][]{planck13-20}. 
We refer to \cite{catalog2} for an exhaustive discussion on how $Y_{SZ}$ is recovered 
from the integrated Comptonisation $Y_{5 R_{500}}$, which represents a nearly unbiased proxy for the total SZ flux 
within a cylinder of aperture radius $5 R_{500}$ and needs to be corrected for the degeneracy induced from 
the signal-size correlation and underlying variations of the pressure profile of reference inducing extra scatter and 
bias in the extrapolation \citep[see Sections~5.1-5.3 and Fig.~16 in ][]{catalog2}.

A second issue is that the gas fraction $f_{nc}$ in Eq.~\ref{eq:ysz_t} 
refers to the clumping-free value, i.e. a gas mass fraction that does not depend on the gas clumping
described by the factor $C$ that appears in Eq.~\ref{eq:scalaw} above. 
By combining now the scaling relations based on SZ data and those using X-ray quantities, and 
modelling the gas clumping as $C = C_0 \,  \mathcal{T}^{C_1}$, we can write
$f_g = C^{0.5} f_{nc} = C_0^{0.5} f_{nc,0} \,  \mathcal{T}^{f_{nc,1} +0.5 C_1} = f_0 \,  \mathcal{T}^{f_1}$, where
$f_{nc,0}$ and $f_{nc,1}$ indicate the normalisation and slope, respectively, 
of the gas fraction after the correction for the clumping.
After simple calculations we obtain
\begin{equation}
\begin{cases}
C_0^{0.5} = f_0 \, t_0^{5/2} \, (N_{YT}/k_{SZ})^{-1} \\
C_1 = 2 f_1 -2 S_{YT} +5 +5 t_1 \\
f_{nc,0} = f_0 \, C_0^{-0.5} \\
f_{nc,1} = f_1 -0.5 C_1
\end{cases},
\label{eq:c0}
\end{equation}
and proceed with the estimates of the gas clumping and the corrected gas fraction
using the values of $f_T$ and $f_g$ quoted in Equations~\ref{eq:mod} and \ref{eq:esz}.

\begin{figure}
\includegraphics[page=12,trim=30 40 60 240,clip,width=\hsize]{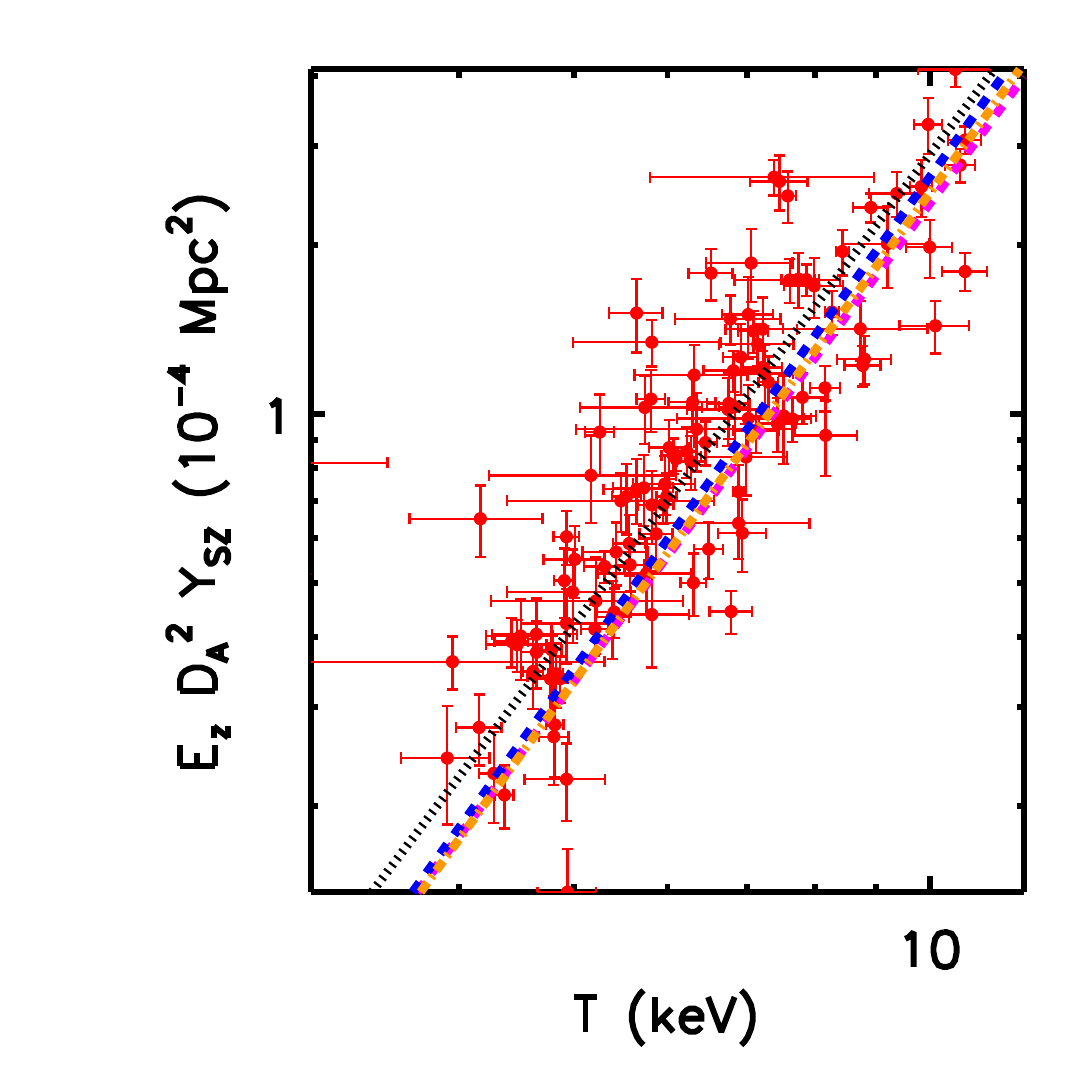}
\caption{$Y_{SZ} - T$ relation for the ESZ sample with the best-fit results from a linear fit in the logarithmic space.
The fit on the observed data is performed using {\tt LIRA} 
(black dotted line).
% both {\tt LINMIX\_ERR} (black dashed line) and {\tt LIRA} (black dot-dashed line).
The predictions from the semi-analytic model overlap (magenta dashed line: D14 model; blue dashed line: B13 model). 
The dot-dashed orange line is obtained from the propagation of the scaling in the E15 formalism (see Eq.~\ref{eq:ysz_t}).
%(Right) Expected changes in the normalization $N_{YT}$ (red solid line) and slope $S_{YT}$ (green dashed line) in the $Y_{SZ} - T$ relation 
%as function of the bias $b$. Normalization and slope are divided by the best-fit values from the ESZ sample.
%The dotted lines indicate the errors propagated from the best-fit values.
%The dependence of the gas fraction on $b$ is also shown (dot-dashed blue line).
} \label{fig:ysz}
\end{figure}

We present the results of our analysis in Fig.~\ref{fig:ysz}.
By fitting a linear relation in the log space on the sample of ten objects simulated with our semi-analytic model to mimic 
the observed distribution in mass and redshift of the ESZ sample, we measure 
$(N_{YT}, S_{YT}) = (0.33, 2.84)$ when a D14 $c$-$M$-$z$ relation is assumed,
and $(0.35, 2.92)$ for the B13 relation.
Using the relations imposed by our universal model (see Eq.~\ref{eq:c0}), we constrain the gas clumping 
to be $(C_0, C_1) = (1.00, 0.12)$ and $(1.04, 0.17)$ for the D14 and B13 model, respectively.
Considering that our model does not include any gas clumping, we can consider these values 
as indicators of the systematic uncertainties (on the order of a few per cent) that affect our reconstruction 
of the ICM properties.

When the same analysis is applied to the ESZ sample, we estimate $(N_{YT}, S_{YT}) = (0.42 \pm 0.03, 2.79 \pm 0.18)$.
These results are  consistent with the best-fit constraints obtained for 62 objects listed in a \planck\ Early Results 
work \citep[][redoing the fit, we measure a normalisation in the adopted units of $0.37 \pm 0.01$  
and slope of $2.97 \pm 0.18$]{planck11.xi}.
Adopting as reference the best-fit values for the ESZ sample, we investigate how we can reconcile the observed differences
with our predictions.
We note that the normalisation $N_{YT}$ depends on the gas fraction and $f_T$.
As discussed in Sect.~\ref{sect:bias}, any hydrostatic bias induces lower values of $f_g$ and $f_T$, 
decreasing the normalisation and enlarging the tension with the observed constraints.
By combining the constraints on $f_g$ and $f_T$ with the relations in Eq.~\ref{eq:c0},
we conclude that the ESZ dataset is consistent with a gas clumping $C_0 < 1.4$ (at the 1$\sigma$ confidence level),
 in close agreement with the expected values within $R_{500}$ from hydrodynamical simulations 
\citep[see e.g. ][]{nagai+11,roncarelli+13,vazza+13,eckert+15}.
From the best-fit results, we can also  constrain $C_1 \sim 1.0 (\pm 0.5)$, which represents a further contribution 
to the dependence on $T$ of the gas fraction, explaining the steeper dependence quoted in Eq.~\ref{eq:esz}
with respect to the predictions from the model.

\section{Discussion}
\label{sect:discussion}

Our model combines a universal profile for the gas pressure with the present knowledge on the distribution of the dark
matter in galaxy clusters to calibrate properly any deviations from the standard self-similar scenario.
In particular, three quantities are introduced to account for these deviations: 
a temperature-dependent gas mass fraction ,$f_g$; a temperature-dependent ratio between the temperature at given radius and
its global value, $f_T$; a hydrostatic bias, $b$.
To explain self-consistently the scaling relations observed in our model and those recovered in the ESZ sample,
we require a significant dependence on $T$ of the gas fraction, and a milder dependence of $f_T$.

As numerical simulations suggest, and also analyses of the observed gas density distribution as a function of the measured temperature, 
more massive (hotter) systems tend to have a relatively higher gas density in the cores than groups 
\citep[see e.g.][]{croston08,eckert+16}, where feedback from central AGNs can efficiently contrast 
the attraction from gravity and push gas particles beyond $R_{500}$.
The net effect is a reduction of the gas mass fraction in objects at lower masses.
This variation of the gas mass fraction with the gas temperature (or total mass) is well 
documented in  past works, mostly based on X-ray selected samples \citep[e.g.][]{pratt09,eckert13fgas,lovisari15,ettori15,eckert+16},
with constraints on normalisations and slopes that are quite similar, but not identical, to the values we obtain in our study 
and summarise in Equations~\ref{eq:esz}.
For example, by converting to our units, \cite{pratt09} estimate $f_g  =0.107 \times (T/5 {\rm keV})^{0.36}$;
\cite{lovisari15} measure $f_g  =0.106 \times (T/5 {\rm keV})^{0.32}$; 
\cite{ettori15} obtains $f_g  =0.107 \times (T/5 {\rm keV})^{0.33}$; and 
\cite{eckert+16} obtain $f_g  =0.079\times (T/5 {\rm keV})^{0.35}$, which is 
on the lower side probably due to a bias in the weak-lensing mass measurements \citep[see e.g.][]{umetsu20}.
Using the estimates within $R_{500}$ of the gas mass, total mass, and temperature obtained for the ESZ sample, 
we can directly fit the gas fraction--temperature relation and obtain 
$f_g = 0.123 (\pm 0.002) \times (T/5 {\rm keV})^{0.39 (\pm 0.05)}$,
in agreement with our indirect calculations.
However, these values of the gas fraction at a given temperature are
15-20\%\  higher than those from the literature cited above.
As discussed in \cite{lovisari20}, part of this  offset can be attributed to the nature of the ESZ sample, 
containing a larger number of disturbed clusters for which higher gas fractions are measured 
\citep[see also][]{eckert13fgas} as a cumulative effect of gas clumpiness and inhomogeneities 
(inducing higher gas mass),  larger non-thermal contribution to the total pressure (biasing low the total mass), 
and possible violation of the hydrostatic equilibrium. 
It is indeed known that SZ selected samples tend to have 
a larger contribution of dynamically disturbed clusters than X-ray selected ones, 
as also confirmed for the ESZ sample \citep[see e.g.][]{lovisari17}.

We can evaluate the impact of having more relaxed objects on our relations.
We repeat the analysis by assuming a pressure profile for cool-core systems, with
$(P_0, c_{500}, \Gamma, A, B) = (11.82, 0.60, 0.31, 0.76, 6.58)$ \citep[see][]{planck13}, 
and using the $c$-$M$-$z$ relation for relaxed systems in \cite{bha13}.
Still assuming no hydrostatic bias ($b=0$), higher mass concentrations associated with relaxed objects
produce higher normalisations of $f_g$ by 5-10\%. 
To compensate for this rise affecting more X-ray selected samples, and to further reduce their estimates of $f_g$, 
we are forced to require the presence of a hydrostatic bias of about 0.2.
As we discuss in Sect.~\ref{sect:bias}, the presence of any hydrostatic bias propagates to all the derived quantities,
lowering $f_g$ for a given initial set of $(M_{500}, z)$. 
A bias of 0.2 causes a reduction of 15\% or more in the normalisation of $f_g$, allowing us, by compensating for the increase 
due to a larger contribution of relaxed clusters, to match the published results based on X-ray selected samples.
A similar (or even higher) bias should also be present   in the ESZ sample due to the   larger contribution of disturbed systems.
The presence of more disturbed systems should induce   larger inhomogeneities in the gas distribution,
which would bias high the reconstructed gas density and the corresponding gas mass fraction \citep[e.g.][]{roncarelli+13}, so that
any correction for it would move $f_g$ to lower values.

On the other hand, the results on $f_g$ in the ESZ sample matches the predictions of our semi-analytic model with $b=0$. 
It is worth noting that the adopted input models describe   the ESZ sample best 
because we consider the average pressure profile recovered for \planck\ selected systems and use the $c$-$M$-$z$ relation estimated 
for the entire collection of haloes (i.e. also including  more disturbed systems). 
So, it seems that a sort of conspiracy acts to mimic $f_g$ with $b=0$, although we are fairly confident that the ESZ sample contains a 
larger number of disturbed objects   than in X-ray selected samples.

\begin{figure}
\includegraphics[page=13,trim=30 40 60 240,clip,width=\hsize]{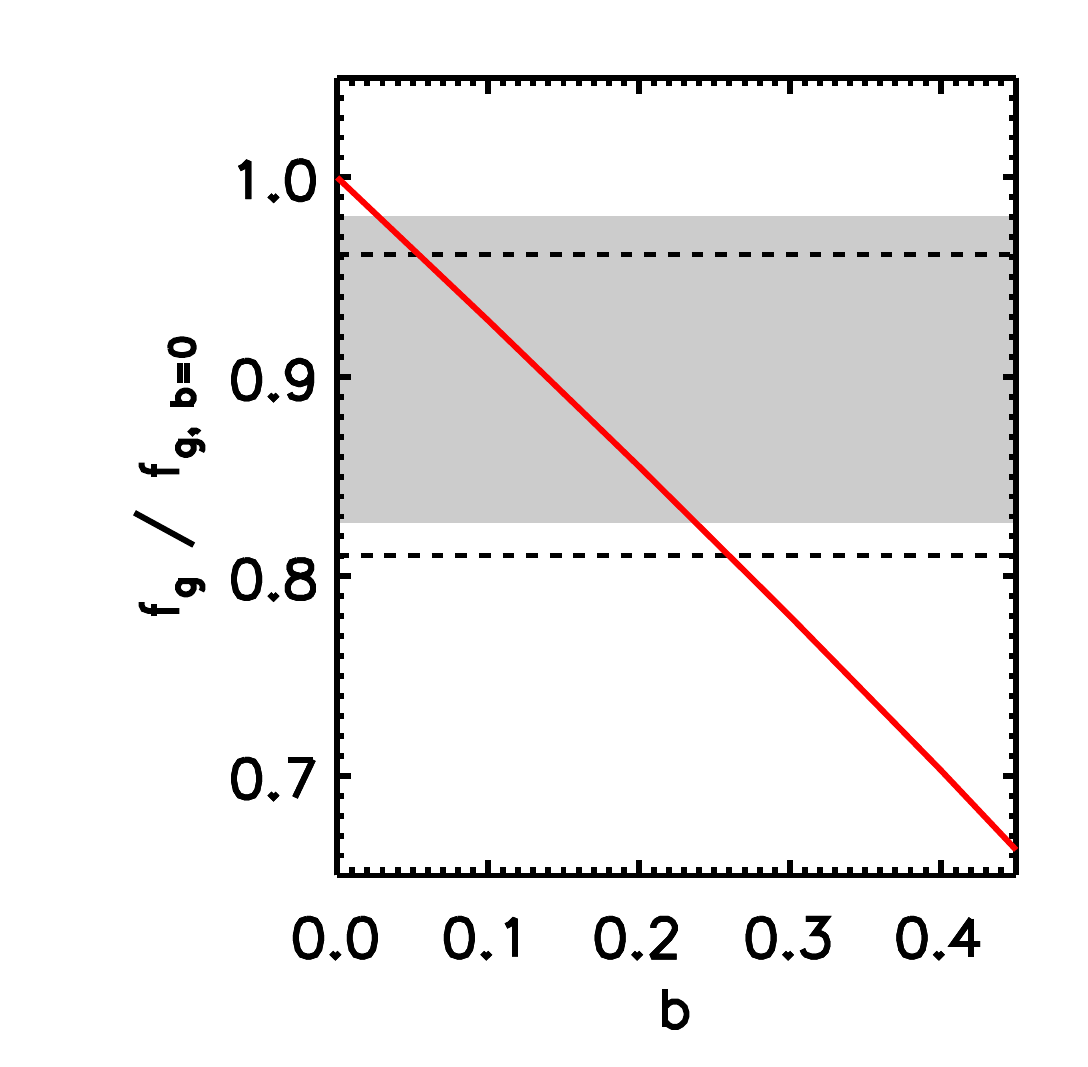}
\caption{Dependence of the normalisation of $f_g$ upon the bias $b$ (red line; see Eq.~\ref{eq:fg_b}).
The shaded region (and the region between dashed lines) represents the quantity $f_{nc} / f_{g, b=0}$, where $f_{nc} = f_g / C_0^{0.5}$,
with $f_g  \approx 0.123$, $f_{g, b=0} = 0.124 \, (0.126)$ and $1< C_0 < 1.4$ being the $1\sigma$ upper limit on the level of gas clumping.
} \label{fig:fg_b}
\end{figure}

As we discuss in Sect.~\ref{sect:bias}, the hydrostatic bias impacts all the integrated quantities.
We evaluate how it affects the normalisation of $f_g$ (see also Eq.~\ref{eq:mod_bias}), and model it as
\begin{equation}
\frac{f_g}{f_{g, b=0}} =  \alpha_0 \, (1-b) +(1-\alpha_0) \, (1-b)^{\alpha_1}
\label{eq:fg_b}
\end{equation}
with $(\alpha_0, \alpha_1) = (0.48, 0.44)$. Assuming that $f_g  = C_0^{0.5} f_{g, c} \approx 0.123$ is the observed value including the gas clumping, and
$f_{g, b=0} = 0.124 \, (0.126)$ is set from our models (see end of Sect.~\ref{sect:esz}), by adopting the upper limit on the gas clumping 
obtained from the $Y_{SZ}-T$ relation (see Sect.~\ref{sect:yt}), we conclude that the level of hydrostatic bias allowed in the ESZ sample 
has to be below 0.24 (0.26) at  the $1 \sigma$ level for the D14 (B13) model (see Fig.~\ref{fig:fg_b}).

The evidence that the ESZ sample includes disturbed systems might then account for the measured $f_g$ 
by a combination of the presence of hydrostatic bias ($b>0$) and gas clumping ($C>1$) induced 
by large gas inhomogeneities.

The other quantity we  introduced, $f_T = T(R_{500}) / T_{500} \times T_{500} / T_{\rm spec} = T(R_{500}) / T_{\rm spec}$,
has never been investigated before and comes from our request to build a proper normalisation for the 
$M-T$ relation, where the hydrostatic mass depends on the measurement of the temperature at a given radius ($T(R_{500}) $).
As we show in Fig.~\ref{fig:model}, $f_T$ is expected to increase mildly in the hotter systems, with a global
effect that might account for some of the deviations observed in the standard self-similar scenario.
Moreover, the normalisation decreases with increasing hydrostatic bias $b$ (see Fig.~\ref{fig:par}).

\section{Conclusions}

The gravitational force aggregates matter onto galaxy clusters, driving them to the virialisation and regulating 
the distribution and the energetic budget of the accreted baryons and the emergent observational properties of the ICM.
We show how a universal pressure profile of the ICM, combined with a halo 
mass -- concentration -- redshift relation (either the model in D14 or in B13) and the hydrostatic equilibrium equation, 
allows  the radial profiles of the thermodynamic quantities to be reconstructed. 
Once integrated over some typical scale (e.g. $R_{500}$), these quantities produces observables 
(such as gas mass, temperature, luminosity, total mass, and SZ Compton parameter), which satisfy 
universal scaling laws that are the simple combination of self-similar relations, 
regulated from the mass and redshift of the halo, and further dependences of the gas temperature measured 
at the radius of reference, $T(R_{500})$, and of the gas mass fraction, $f_g = C^{0.5} f_{nc}$, 
on the observed spectroscopic measurement, $T_{\rm spec, 0.15-1 R_{500}}$.

We  calibrated these dependences both within our framework and using one of the largest samples of X-ray luminous galaxy clusters 
that has been homogeneously  analysed, the ESZ sample \citep{lovisari17,lovisari20}.
We demonstrate that self-similar scaling laws hold between the integrated X-ray observables, 
once $f_T = T(R_{500}) / T_{500} \times T_{500} / T_{\rm spec}$ and 
$f_g = C^{0.5} \, f_{nc}$ are allowed to depend on $T \equiv T_{\rm spec}$  as 
$f_T = t_0  \mathcal{T}^{t_1}$ and $f_g = f_0  \mathcal{T}^{f_1}$ 
(see also Table~\ref{tab:mod} for a description of these quantities)
with the following constraints on the parameters obtained 
from our semi-analytic model (values for the B13 $c$-$M$-$z$ relation in parentheses):
\begin{equation}
\begin{cases} 
f_T = 0.687 \, (0.711) \times (T/5 {\rm keV})^{0.07 (0.10)} \\
f_g = 0.124 \, (0.128) \times (T/5 {\rm keV})^{0.23 (0.25)}
\end{cases}.
\label{eq:final}
\end{equation}

In the ESZ sample, by propagating these dependences to the SZ signal, 
and interpreting any mismatch between the best-fit results of the observational data 
and the predictions from our semi-analytic and theoretical models as a measure of the gas clumping, 
we estimate 
\begin{equation}
\begin{cases}
f_{T, ESZ} = 0.697 (\pm 0.103) \times (T/5 {\rm keV})^{0.15 (\pm 0.06)} \\
f_{g, ESZ} =  0.121 (\pm 0.045) \times (T/5 {\rm keV})^{0.45 (\pm 0.09)} \\
C_{ESZ} = (<1.4)  \times (T/5 {\rm keV})^{1.0 (\pm 0.5)}
\end{cases}.
\label{eq:final_esz}
\end{equation}
This level of gas clumping within $R_{500}$ is consistent with the value measured in
hydrodynamical simulations, and supports the evidence that SZ selected samples, like the ESZ one, tend to have 
a representative contribution of dynamically disturbed clusters \citep[see e.g.][]{lovisari17}
with a relative significant presence of gas inhomogeneities, in particular at large scales 
\citep[see e.g.][]{roncarelli+13,vazza+13}.
This upper limit on the gas clumping implies a level of hydrostatic bias $b$ 
(see Eq.~\ref{eq:fg_b} and Fig.~\ref{fig:fg_b}) below 25\%, on average.
 
We conclude that our semi-analytic model reproduces well the observed properties of galaxy clusters, 
both resolved spatially and as integrated quantities. 
By providing the calibration of this physically motivated model in terms of the standard self-similar scaling relations,
modified by three components that are able to account completely for the observed deviations, 
we can also deduce some other interesting properties.

\begin{table}[ht]
\centering 
\caption{Dependences of the characteristic physical scales on the temperature and mass 
in the universal model. These scaling laws modify the dependences shown in Table~\ref{tab:delta}.}
\begin{tabular}{ccc}
\hline
Quantity  &  $f(T)$  & $f(M)$ \\
\hline 
\rule{0pt}{2.5ex} $T_{\Delta}$ & $T^{1+t_1}$ &  $M^{2/3 / (1+t_1)}$ \\
\rule{0pt}{2.5ex} $\bar{n}_{\rm e}$ & $T^{f_1}$ & $M^{2/3 \, f_1/(1+t_1)}$ \\
\rule{0pt}{2.5ex} $P_{\Delta}$ & $T^{1 + t_1 + f_1}$ & $M^{2/3 \, (1 + t_1 +f_1)/(1+t_1)}$  \\
\rule{0pt}{2.5ex} $K_{\Delta}$ & $T^{1 + t_1 -2/3 f_1}$ & $M^{2/3 \, (1 + t_1 -2/3 f_1)/(1+t_1)}$ \\
\hline
\end{tabular}
\label{tab:delta_univ}
\end{table}

For instance, these explicit forms of the dependences on the observed gas temperature allow us to modify accordingly the characteristic 
physical quantities that renormalises the observed profiles. In Table~\ref{tab:delta_univ} we quote how the dependences 
implied by the universal model propagate and modify some of these quantities.
For example, $P_{500}$, which in a self-similar model scales as $M_{500}^{2/3}$,  is expected to scale 
(in parentheses, the values for the B13 model) as $T^{1.29 (1.35)}$ or $M_{500}^{0.81 (0.82)}$, 
which is consistent with the rescaling of $M_{500}^{2/3 +0.12} \approx M_{500}^{0.79}$ 
suggested in \cite{arnaud10} \citep[and adopted in][]{planck13}.

Moreover, we can write the baryonic depletion parameter $Y$ with its explicit dependence on the gas temperature (or mass) as
\begin{equation}
Y \frac{\Omega_b}{\Omega_m} = f_g + f_s
,\end{equation}
where $f_s$ represents the stellar mass fraction \citep[$\sim 0.015$; see e.g.][]{eckert19}, 
and the ratio of the cosmic baryon density to matter density parameters, 
$\Omega_b / \Omega_m$, is equal to 0.157 \citep{planck+16}.
From the results of our semi-analytic model with a D14 (B13) $c$-$M$-$z$ relation, we predict an average value of 
$Y = 0.787 (0.813) \, (T/5 {\rm keV})^{0.23 (0.25)}+0.095$.

In the same framework, we show that any calibration of the $M_{tot}-T$ relation with mass measurements that do not rely on the hydrostatic 
equilibrium equation can be used to constrain the hydrostatic bias $b$, preserving the use of the self-similar relations  in this case as well.
On the other hand, by controlling the bias $b$ we can make predictions on the expected variations in the observed properties and, for example,
explain the published relations between gas mass fraction and temperature (see Sect.~\ref{sect:discussion}) 
by requiring a hydrostatic bias $b \ga 0.2$, as illustrated in Sect.~\ref{sect:bias}.

The application of this model to a accurately selected sample of a large number ($\sim100$) 
of objects analysed homogeneously in their X-ray and lensing signal out to $R_{500}$ and beyond 
as the one that will  soon be available for the {\it XMM-Newton Heritage Galaxy Cluster Project}\footnote{\url{http://xmm-heritage.oas.inaf.it/}}
will allow us to extend the calibration and our understanding of the physical processes that regulate 
the interplay of baryons and dark matter in galaxy clusters.

Overall, our study demonstrates that the observed properties, both spatially resolved and  integrated
values, of X-ray luminous galaxy clusters are understood well.
This represents a further step in the process to standardise their observables on the basis of a physically motivated model,
not only to fully appreciate the phenomena that shape the distribution of baryons and regulate their energetic budget, but
also to control biases that could affect their use as cosmological proxies.

\begin{acknowledgements}
We thank the anonymous referee for insightful comments that helped in improving the presentation of the work.
We acknowledge financial contribution from the contracts ASI-INAF Athena 2015-046-R.0, ASI-INAF Athena 2019-27-HH.0,
``Attivit\`a di Studio per la comunit\`a scientifica di Astrofisica delle Alte Energie e Fisica Astroparticellare''
(Accordo Attuativo ASI-INAF n. 2017-14-H.0),
and from INAF ``Call per interventi aggiuntivi a sostegno della ricerca di main stream di INAF''.
This research has made use of the SZ-Cluster Database operated by the Integrated Data and Operation Center (IDOC) at the Institut d'Astrophysique Spatiale (IAS) under contract with CNES and CNRS.
\end{acknowledgements}

\bibliographystyle{aa} 
\bibliography{e15l20} 

\begin{appendix}
\section{Derivations of $k_M$ and $k_L$}
\label{app:km_kl}

In equation~\ref{eq:scalaw} we introduce some constants that appear in the normalisations of the scaling relations $M_{tot}-T$ and $L-T$, namely
$k_M$ and $k_L$.
We present here the steps that lead to their definitions.

\begin{figure}
\includegraphics[page=7,trim=0 40 0 230,clip,width=\hsize]{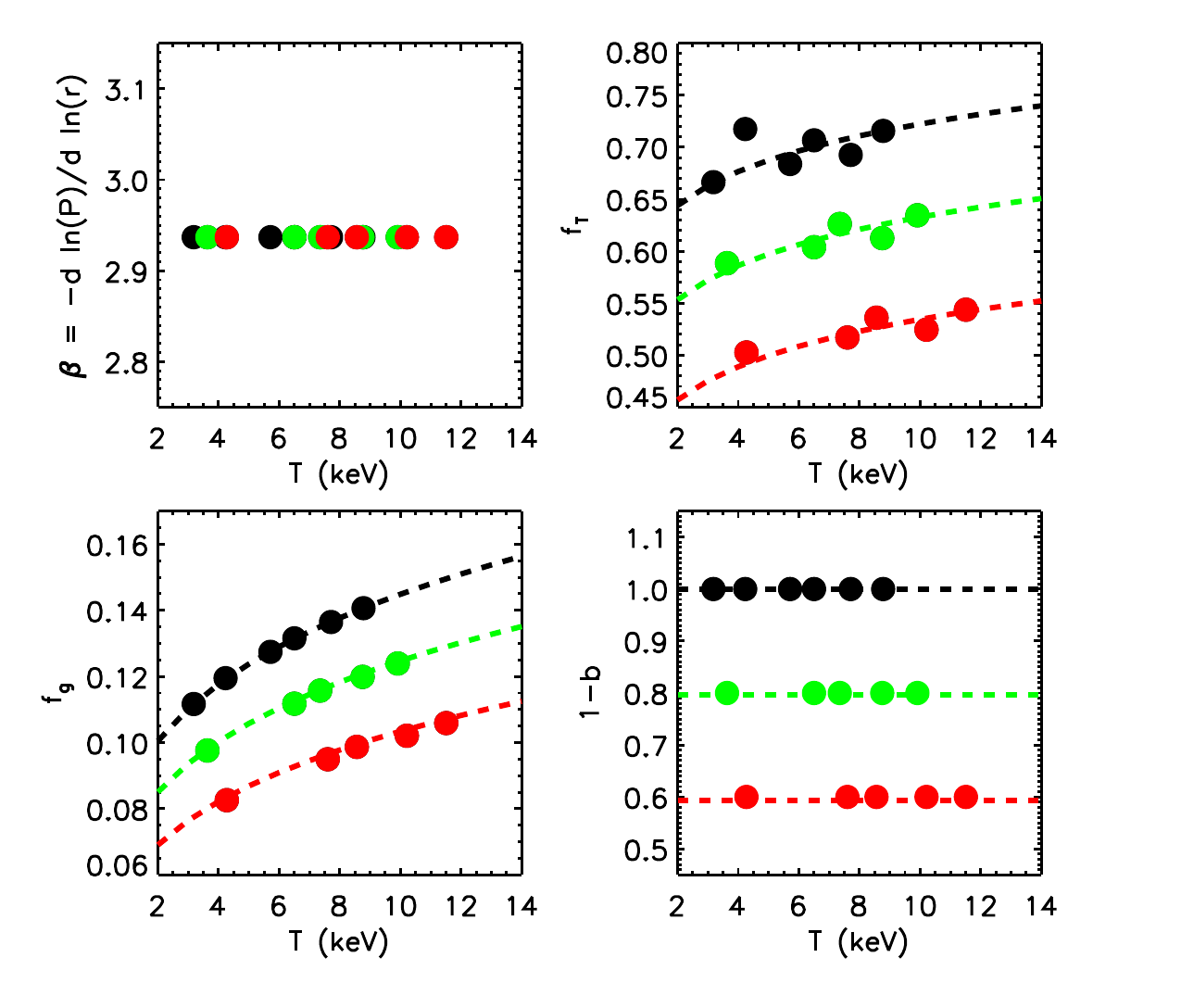} 
\caption{
Distribution of the parameters $\beta$, $f_T$, $f_g$, and $1-b$ as a function of $T$ and colour-coded according to the assumed 
hydrostatic bias (black: $b=0$; green: $b=0.2$; red: $b=0.4$).
The dashed lines in the panel at the bottom right are recovered through the process 
described in Equation~\ref{eq:param}.
Points represent quantities derived for the following input values for $(M_{500}, z)$: 
($8 \times 10^{14} M_{\odot}, 0.05$), ($5 \times 10^{14} M_{\odot}, 0.05$), ($2 \times 10^{14} M_{\odot}, 0.05$),
($8 \times 10^{14} M_{\odot}, 0.5$), ($5 \times 10^{14} M_{\odot}, 0.5$), with and without a bias $b$;
a further case ($2 \times 10^{14} M_{\odot}, 1$) with $b=0$ is considered.
} \label{fig:par}
\end{figure}

The hydrostatic mass at the given overdensity $\Delta$ is defined as
\begin{equation}
M_{\rm HE} = M_{tot} \, (1-b) = -\frac{f_T \, T_{\rm spec} R_{\Delta}}{\mu m_a G} \frac{d \log P}{d \log r},
\label{eq:mhyd}
\end{equation}
where we have defined the quantity $f_T = T(R_{\Delta}) /  T_{\rm spec}$ that relates the gas temperature at $R_{\Delta}$ 
with the observed global value $T_{\rm spec}$.
By definition, the same mass is equal to $M_{\rm HE} = 4/3 \pi \rho_{c,z} \Delta R_{\Delta}^3$, where $\rho_{\rm c,z} = 3 H_z^2 / (8 \pi G)$.
By inverting the latter equation, we obtain $R_{\Delta}^3 = 2 G M_{\rm HE} / (\Delta H_z^2)$ that we use to replace $R_{\Delta}$ in 
eq.~\ref{eq:mhyd}. After this substitution, we solve for $M_{tot}$ and obtain
\begin{equation}
M_{tot} = \left( \frac{2 G}{H_z^2 \Delta} \right)^{1/2} \left( \frac{f_T \, \beta}{\mu m_a G} \right)^{3/2}  (1-b)^{-1} \, T_{\rm spec}^{3/2},
\end{equation}
where we define $\beta$ as the logarithmic slope of the pressure profile at $R_{\Delta}$ ($\beta = -d \log P / d \log r$ 
\footnote{Differently from what we describe in \cite{ettori15}, here the logarithmic slope $\beta$ of the pressure profile at $R_{500}$ 
does not depend  on the mass because the further exponent $\alpha_1$ is set to 0.}).
Finally, collecting all the constants on the right side of the equation, and recalling that $H_z = E_z H_0$, we can write
\begin{equation}
E_z M_{tot} = \left( \frac{f_T \, \beta}{\mu m_a} \right)^{3/2}  \frac{2^{1/2} (1-b)^{-1}}{H_0 G \Delta^{1/2}} \, T_{\rm spec}^{3/2},
\end{equation}
which allows us to define the constant $k_M$ as the normalisation of the relation
\begin{equation}
E_z (M_{tot}/M_0) = k_M \, f_T^{3/2} \, (1-b)^{-1} \, (T_{\rm spec}/T_0)^{3/2}
\label{eq:mt}
,\end{equation}
where
\begin{equation}
k_M = \left( \frac{ \bar{f_T} \bar{\beta} }{\mu m_a}\right)^{3/2} \frac{2^{1/2}}{H_0 G \Delta^{1/2}}  \frac{T_0^{3/2}}{M_0}.
\end{equation}
Using $\Delta=500$ and the median values $\bar{\beta} = 2.94$ (identical for all the systems being the same as the pressure profile
adopted) and $\bar{f_T} = 0.71$ (with values in the range of $-$6\%, +2\% for $b=0$; see Fig.~\ref{fig:par}), and
the adopted pivot values $T_0$ and $M_0$, we measure $k_M = 0.883$.

Consider now the $L-T$ relation.
By definition, the X-ray luminosity is equal to the integral over the volume $V$ of interest of the X-ray emissivity 
$\epsilon = n_e n_p \Lambda(T,Z)$:
\begin{equation}
L = \int \epsilon \, dV.
\end{equation}
A selection of the X-ray band where the luminosity is measured will affect the estimate of the cooling function $\Lambda(T,Z), $ which depends 
on the gas temperature and metallicity.
For a bolometric luminosity we can write  $\epsilon= c_f \, n_e^2 \, T_{keV}^{1/2}$, with
$c_f = 1.02 \times 10^{-23}$ erg/s/cm$^3$ (see  discussion in sect.~2 of E15), and write
\begin{equation}
L = c_f \, \frac{f_L}{\mu_e^2 m_a^2} \frac{M_g^2}{V} \,  T_{\rm spec}^{1/2} 
  = c_f \, \frac{f_L}{\mu_e^2 m_a^2} \, f_g^2 \, \frac{M_{\rm HE}^2}{V} \,  T_{\rm spec}^{1/2},
\end{equation}
where we have defined $f_L = \int{\rho_{\rm gas}^2 dV} / (\int{\rho_{\rm gas} dV})^2 \, V$ as the correction needed to consider the gas mass 
($M_g = \int{\rho_{\rm gas} dV}$) instead of the emission integral ($\int{n_e^2 dV}$), with $\rho_{\rm gas} = \mu_e m_a n_e$, for  scaling purposes.
Here the integrals are performed over a spherical volume between 0 and $R_{500}$.
Using the above relation between total mass and temperature and total mass and volume, we can do the final step and write
\begin{equation}
L = c_f \, \frac{f_L}{\mu_e^2 m_a^2} \, f_g^2  \, f_T^{3/2} \, \Delta \rho_{cz}  \frac{k_M M_0}{E_z T_0^{3/2}}  \,  T_{\rm spec}^2 
,\end{equation}
which can be expressed in the form
\begin{equation}
E_z^{-1} (L/L_0) = k_L \, f_g^2 \, f_T^{3/2} \, (T_{\rm spec}/T_0)^2 
\end{equation}
with 
\begin{equation}
k_L = k_M \frac{\bar{f_L} c_f f_{g,0}^2}{\mu_e^2 m_a^2} \frac{3 \Delta H_0^2}{8 \pi G} \frac{M_0 T_0^{1/2}}{L_0}.
\end{equation}
Using the median value of $f_L$, $\bar{f_L} = 1.90$ (with values in the range --8\%, +2\%), $f_{g,0}=0.1$,
and the adopted values of $\Delta$, $T_0$, $M_0$, and $L_0$, we measure $k_L = 0.930$.

\section{Derivation of $k_{SZ}$}
\label{app:ksz}

In equation~\ref{eq:ysz_t} we introduce the constant $k_{SZ}$ in the normalisation of the $Y-T$ relation:
\begin{equation}
\frac{E_z Y_{SZ} D_A^2}{10^{-4} {\rm Mpc}^2} = k_{SZ} \, f_T^{5/2} \, f_{nc} \, \mathcal{T}^{5/2}.
\end{equation}
Here $f_T$ and $f_g$ are  the values normalised to the median value we observe in our model $\bar{f_T} = 0.71$ (see Fig.~\ref{fig:par})
and 0.1, respectively, and $f_{nc}$ is the clumping-free gas mass fraction that is related to the X-ray-measured quantity $f_g$ through the relation
$f_g = C^{0.5} \, f_{nc}$, where  $C = <n_{\rm gas}^2> / <n_{\rm gas}>^2$ is the clumping factor.

To define $k_{SZ}$, we proceed from the definition of $Y_{SZ} D_A^2 = (\sigma_T/ m_e c^2) \int P_e dV$, where $D_A$ is the angular diameter distance to the cluster, 
$\sigma_T = 8 \pi/3 ( e^2 / m_e c^2 )^2 = 6.65 \times 10^{-25}$ cm$^2$ is the Thomson cross section, and
$\int P_e dV = \int  n_e T dV \approx f_T \, T_{\rm spec} \, M_g C^{-0.5} / (\mu_e m_a) = f_g \, C^{-0.5} \, f_T \, T_{\rm spec} M_{\rm HE} / (\mu_e m_a)$.

By substituting $M_{\rm HE}$ from eq.~\ref{eq:mt}, we can write
\begin{equation}
Y_{SZ} D_A^2 = \frac{\sigma_T}{m_e c^2}  \frac{f_{nc}}{\mu_e m_a} \, f_T^{5/2} \frac{k_M M_0 T_0}{E_z} \mathcal{T}^{5/2},
\end{equation}
which allows us to define the constant
\begin{align}
k_{SZ} & = \frac{\sigma_T}{m_e c^2} c_P \frac{2^{1/2}\, f_{g,0} \, {\bar{\beta}}^{3/2} \, \bar{f_T}^{5/2}}{G H_0 \Delta^{1/2} \mu_e \mu^{3/2} m_a^{5/2}}
% \left( \frac{\bar{\beta}}{\mu m_a}\right)^{3/2}  
\frac{T_0^{5/2}}{Y_0} = 0.296, % \nonumber \\
% & T_0^{5/2} / (10^{-4} {\rm Mpc}^2) = 0.296
\label{eq:ksz}
\end{align}
where $f_{g,0} = 0.1$, $\bar{\beta} = 2.94$, $\bar{f_T}=0.71$, $T_0 = 5$ keV, $Y_0 =  10^{-4} {\rm Mpc}^2$, and 
$c_P = \int{P_e dV} \times (\mu_e m_a) / (M_g  f_T T_{\rm spec})$
has a typical value of 1.36 (dimensionless) with variations of $\pm1$\% in the range of $T$ and redshift studied here.

It should be noted that using  $Y_X = M_g \times T$, which is the X-ray analogue of the integrated Compton parameter 
introduced by \cite{kravtsov06}, we can write $c_P = Y_{SZ} / (c_{XSZ} Y_X f_T)$, where $c_{XSZ} = \sigma_T/ (m_e c^2 \, \mu_e m_a) 
\approx 1.40 \times 10^{-19}$ Mpc$^2$ $M_{\odot}^{-1}$ keV$^{-1}$. Therefore, we predict a ratio $C = Y_{SZ} / (c_{XSZ} Y_X) = c_P \, f_T$
equal to median values of 0.924 (with D14 model) and 0.927 (B13) with relative changes of $\pm1$\% in the redshift and mass ranges  investigated in this paper,
 in close agreement with the present observational constraints \citep[see e.g.][]{arnaud10}.

\end{appendix}

\end{document}